\documentclass[twocolumn, resetfootnote]{aastex7}
\shorttitle{Active Black Holes from DESI DR1}
\shortauthors{Pucha et al.}

\usepackage{amsmath}
\usepackage{amssymb}
\usepackage{longtable}
\usepackage{subcaption}
\usepackage[table]{xcolor}

\newcommand{\ha}{\textrm{H}\ensuremath{\alpha}}
\newcommand{\hb}{\textrm{H}\ensuremath{\beta}}
\newcommand{\oi}{[\textrm{O}\,\textsc{i}]}

\newcommand{\oiii}{[\textrm{O}\textsc{iii}]}
\newcommand{\nii}{[\textrm{N}\textsc{ii}]}
\newcommand{\sii}{[\textrm{S}\textsc{ii}]}
\newcommand{\halam}{\textrm{H}\ensuremath{\alpha}\ensuremath{\lambda}6563}
\newcommand{\hblam}{\textrm{H}\ensuremath{\beta}\ensuremath{\lambda}4861}

\newcommand{\siilam}{[\textrm{S}\textsc{ii}]\ensuremath{\lambda\lambda}6717,6731}
\newcommand{\niilamlam}{[\textrm{N}\textsc{ii}]\ensuremath{\lambda\lambda}6548,6584}
\newcommand{\oiiilamlam}{[\textrm{O}\textsc{iii}]\ensuremath{\lambda\lambda}4959,5007}
\newcommand{\feiilamlam}{[\textrm{Fe}\textsc{ii}]\ensuremath{\lambda\lambda}4924,5018}

\newcommand{\mbh}{\ensuremath{M_{\rm BH}}}
\newcommand{\mstar}{\ensuremath{M_{\star}}}
\newcommand{\msun}{\ensuremath{M_{\odot}}}
\newcommand{\logmass}{\ensuremath{\log (\mstar/\msun)}}
\newcommand{\logmbh}{\ensuremath{\log (\mbh/\msun)}}
\newcommand{\ergscmsq}{\textrm{ergs s$^{-1}$ cm$^{-2}$}}
\newcommand{\ergsAAcmsq}{\textrm{ergs s$^{-1}$ \AA$^{-1}$ cm$^{-2}$}}
\newcommand{\kms}{\textrm{km s$^{-1}$}}

\begin{document}

\title{A New Record Census of Dwarf AGN and a Bimodal $\mbh - \mstar$ Scaling Relation with DESI DR1}

\correspondingauthor{Ragadeepika~Pucha}\email{raga.pucha@utah.edu}
\suppressAffiliations

\author[0000-0002-4940-3009]{Ragadeepika~Pucha}
\affiliation{Department of Physics and Astronomy, The University of Utah, 115 South 1400 East, Salt Lake City, UT 84112, USA}
\affiliation{Steward Observatory, University of Arizona, 933 N. Cherry Avenue, Tucson, AZ 85721, USA}
\email[]{raga.pucha@utah.edu}

\author[0000-0002-0000-2394]{St\'ephanie~Juneau}
\affiliation{NSF NOIRLab, 950 N. Cherry Ave., Tucson, AZ 85719, USA}
\email[]{}

\author[0000-0003-4440-259X]{M.~Mezcua}
\affiliation{Institute of Space Sciences, ICE-CSIC, Campus UAB, Carrer de Can Magrans s/n, 08913 Bellaterra, Barcelona, Spain}
\affiliation{Institut d’Estudis Espacials de Catalunya (IEEC), c/ Esteve Terradas 1, Edifici RDIT, Campus PMT-UPC, 08860 Castelldefels, Spain}
\email[]{}

\author[0000-0002-4928-4003]{Arjun~Dey}
\affiliation{NSF NOIRLab, 950 N. Cherry Ave., Tucson, AZ 85719, USA}
\email[]{}

\author[0000-0002-1200-0820]{Y.-Y.~Mao}
\affiliation{Department of Physics and Astronomy, The University of Utah, 115 South 1400 East, Salt Lake City, UT 84112, USA}
\email[]{}

\author[0000-0002-5896-6313]{D.~M.~Alexander}
\affiliation{Centre for Extragalactic Astronomy, Department of Physics, Durham University, South Road, Durham, DH1 3LE, UK}
\affiliation{Institute for Computational Cosmology, Department of Physics, Durham University, South Road, Durham DH1 3LE, UK}
\email[]{}

\author{C.~Circosta}
\affiliation{Department of Physics \& Astronomy, University College London, Gower Street, London, WC1E 6BT, UK}
\email[]{}

\author[0000-0003-1251-532X]{V.~A.~Fawcett}
\affiliation{European Southern Observatory, Karl-Schwarzschild-Strasse 2, 85748 Garching bei München, Germany}
\affiliation{School of Mathematics, Statistics and Physics, Newcastle University, Newcastle upon Tyne NE1 7RU, UK}
\email[]{}

\author[0000-0001-9457-0589]{Wei-Jian Guo}
\affiliation{National Astronomical Observatories, Chinese Academy of Sciences, A20 Datun Road, Chaoyang District, Beijing, 100101, P.~R.~China}
\email[]{}

\author[0000-0002-2733-4559]{J.~Moustakas}
\affiliation{Department of Physics and Astronomy, Siena University, 515 Loudon Road, Loudonville, NY 12211, USA}
\email[]{}

\author[0000-0002-5854-7426]{S.~Panda}
\affiliation{NSF NOIRLab, 950 N. Cherry Ave., Tucson, AZ 85719, USA}
\email[]{}

\author[0000-0002-2949-2155]{M.~Siudek}
\affiliation{Institute of Space Sciences, ICE-CSIC, Campus UAB, Carrer de Can Magrans s/n, 08913 Bellaterra, Barcelona, Spain}
\affiliation{Instituto de Astrof\'{\i}sica de Canarias, C/ V\'{\i}a L\'{a}ctea, s/n, E-38205 La Laguna, Tenerife, Spain}
\email[]{}

\author{Zhefu~Yu}
\affiliation{SLAC National Accelerator Laboratory, 2575 Sand Hill Road, Menlo Park, CA 94025, USA}
\email[]{}

\author{J.~Aguilar}
\affiliation{Lawrence Berkeley National Laboratory, 1 Cyclotron Road, Berkeley, CA 94720, USA}
\email[]{}

\author[0000-0001-6098-7247]{S.~Ahlen}
\affiliation{Department of Physics, Boston University, 590 Commonwealth Avenue, Boston, MA 02215 USA}
\email[]{}

\author[0000-0001-9712-0006]{D.~Bianchi}
\affiliation{Dipartimento di Fisica ``Aldo Pontremoli'', Universit\`a degli Studi di Milano, Via Celoria 16, I-20133 Milano, Italy}
\affiliation{INAF-Osservatorio Astronomico di Brera, Via Brera 28, 20122 Milano, Italy}
\email[]{}

\author{D.~Brooks}
\affiliation{Department of Physics \& Astronomy, University College London, Gower Street, London, WC1E 6BT, UK}
\email[]{}

\author{T.~Claybaugh}
\affiliation{Lawrence Berkeley National Laboratory, 1 Cyclotron Road, Berkeley, CA 94720, USA}
\email[]{}

\author[0000-0002-0553-3805]{K.~S.~Dawson}
\affiliation{Department of Physics and Astronomy, The University of Utah, 115 South 1400 East, Salt Lake City, UT 84112, USA}
\email[]{}

\author[0000-0002-1769-1640]{A.~de la Macorra}
\affiliation{Instituto de F\'{\i}sica, UNAM,  Circuito de la Investigaci\'{o}n Cient\'{\i}fica, Ciudad Universitaria, Cd. de M\'{e}xico  C.~P.~04510,  M\'{e}xico}
\email[]{}

\author{P.~Doel}
\affiliation{Department of Physics \& Astronomy, University College London, Gower Street, London, WC1E 6BT, UK}
\email[]{}

\author[0000-0003-4992-7854]{S.~Ferraro}
\affiliation{Lawrence Berkeley National Laboratory, 1 Cyclotron Road, Berkeley, CA 94720, USA}
\affiliation{University of California, Berkeley, 110 Sproul Hall \#5800 Berkeley, CA 94720, USA}
\email[]{}

\author[0000-0002-3033-7312]{A.~Font-Ribera}
\affiliation{Instituci\'{o} Catalana de Recerca i Estudis Avan\c{c}ats, Passeig de Llu\'{\i}s Companys, 23, 08010 Barcelona, Spain}
\affiliation{Institut de F\'{i}sica d’Altes Energies (IFAE), The Barcelona Institute of Science and Technology, Edifici Cn, Campus UAB, 08193, Bellaterra (Barcelona), Spain}
\email[]{}

\author[0000-0002-2890-3725]{J.~E.~Forero-Romero}
\affiliation{Departamento de F\'isica, Universidad de los Andes, Cra. 1 No. 18A-10, Edificio Ip, CP 111711, Bogot\'a, Colombia}
\affiliation{Observatorio Astron\'omico, Universidad de los Andes, Cra. 1 No. 18A-10, Edificio H, CP 111711 Bogot\'a, Colombia}
\email[]{}

\author[0000-0001-9632-0815]{E.~Gaztañaga}
\affiliation{Institut d'Estudis Espacials de Catalunya (IEEC), c/ Esteve Terradas 1, Edifici RDIT, Campus PMT-UPC, 08860 Castelldefels, Spain}
\affiliation{Institute of Cosmology and Gravitation, University of Portsmouth, Dennis Sciama Building, Portsmouth, PO1 3FX, UK}
\affiliation{Institute of Space Sciences, ICE-CSIC, Campus UAB, Carrer de Can Magrans s/n, 08913 Bellaterra, Barcelona, Spain}
\email[]{}

\author[0000-0003-3142-233X]{Satya~{Gontcho A Gontcho}}
\affiliation{University of Virginia, Department of Astronomy, Charlottesville, VA 22904, USA}
\email[]{}

\author{G.~Gutierrez}
\affiliation{Fermi National Accelerator Laboratory, PO Box 500, Batavia, IL 60510, USA}
\email[]{}

\author[0000-0003-1197-0902]{C.~Hahn}
\affiliation{Department of Astronomy, University of Texas at Austin, 2515 Speedway, TX 78712, USA}
\email[]{}

\author[0000-0002-6550-2023]{K.~Honscheid}
\affiliation{Center for Cosmology and AstroParticle Physics, The Ohio State University, 191 West Woodruff Avenue, Columbus, OH 43210, USA}
\affiliation{Department of Physics, The Ohio State University, 191 West Woodruff Avenue, Columbus, OH 43210, USA}
\affiliation{The Ohio State University, Columbus, 43210 OH, USA}
\email[]{}

\author[0000-0003-0201-5241]{R.~Joyce}
\affiliation{NSF NOIRLab, 950 N. Cherry Ave., Tucson, AZ 85719, USA}
\email[]{}

\author{R.~Kehoe}
\affiliation{Department of Physics, Southern Methodist University, 3215 Daniel Avenue, Dallas, TX 75275, USA}
\email[]{}

\author[0000-0003-3510-7134]{T.~Kisner}
\affiliation{Lawrence Berkeley National Laboratory, 1 Cyclotron Road, Berkeley, CA 94720, USA}
\email[]{}

\author[0000-0001-6356-7424]{A.~Kremin}
\affiliation{Lawrence Berkeley National Laboratory, 1 Cyclotron Road, Berkeley, CA 94720, USA}
\email[]{}

\author[0000-0003-1838-8528]{M.~Landriau}
\affiliation{Lawrence Berkeley National Laboratory, 1 Cyclotron Road, Berkeley, CA 94720, USA}
\email[]{}

\author[0000-0001-7178-8868]{L.~Le~Guillou}
\affiliation{Sorbonne Universit\'{e}, CNRS/IN2P3, Laboratoire de Physique Nucl\'{e}aire et de Hautes Energies (LPNHE), FR-75005 Paris, France}
\email[]{}

\author[0000-0003-4962-8934]{M.~Manera}
\affiliation{Departament de F\'{i}sica, Serra H\'{u}nter, Universitat Aut\`{o}noma de Barcelona, 08193 Bellaterra (Barcelona), Spain}
\affiliation{Institut de F\'{i}sica d’Altes Energies (IFAE), The Barcelona Institute of Science and Technology, Edifici Cn, Campus UAB, 08193, Bellaterra (Barcelona), Spain}
\email[]{}

\author[0000-0002-1125-7384]{A.~Meisner}
\affiliation{NSF NOIRLab, 950 N. Cherry Ave., Tucson, AZ 85719, USA}
\email[]{}

\author{R.~Miquel}
\affiliation{Instituci\'{o} Catalana de Recerca i Estudis Avan\c{c}ats, Passeig de Llu\'{\i}s Companys, 23, 08010 Barcelona, Spain}
\affiliation{Institut de F\'{i}sica d’Altes Energies (IFAE), The Barcelona Institute of Science and Technology, Edifici Cn, Campus UAB, 08193, Bellaterra (Barcelona), Spain}
\email[]{}

\author[0000-0001-9070-3102]{S.~Nadathur}
\affiliation{Institute of Cosmology and Gravitation, University of Portsmouth, Dennis Sciama Building, Portsmouth, PO1 3FX, UK}
\email[]{}

\author[0000-0002-0644-5727]{W.~J.~Percival}
\affiliation{Department of Physics and Astronomy, University of Waterloo, 200 University Ave W, Waterloo, ON N2L 3G1, Canada}
\affiliation{Perimeter Institute for Theoretical Physics, 31 Caroline St. North, Waterloo, ON N2L 2Y5, Canada}
\affiliation{Waterloo Centre for Astrophysics, University of Waterloo, 200 University Ave W, Waterloo, ON N2L 3G1, Canada}
\email[]{}

\author[0000-0001-7145-8674]{F.~Prada}
\affiliation{Instituto de Astrof\'{i}sica de Andaluc\'{i}a (CSIC), Glorieta de la Astronom\'{i}a, s/n, E-18008 Granada, Spain}
\email[]{}

\author[0000-0001-6979-0125]{I.~P\'erez-R\`afols}
\affiliation{Departament de F\'isica, EEBE, Universitat Polit\`ecnica de Catalunya, c/Eduard Maristany 10, 08930 Barcelona, Spain}
\email[]{}

\author{G.~Rossi}
\affiliation{Department of Physics and Astronomy, Sejong University, 209 Neungdong-ro, Gwangjin-gu, Seoul 05006, Republic of Korea}
\email[]{}

\author[0000-0002-9646-8198]{E.~Sanchez}
\affiliation{CIEMAT, Avenida Complutense 40, E-28040 Madrid, Spain}
\email[]{}

\author{D.~Schlegel}
\affiliation{Lawrence Berkeley National Laboratory, 1 Cyclotron Road, Berkeley, CA 94720, USA}
\email[]{}

\author{M.~Schubnell}
\affiliation{Department of Physics, University of Michigan, 450 Church Street, Ann Arbor, MI 48109, USA}
\affiliation{University of Michigan, 500 S. State Street, Ann Arbor, MI 48109, USA}
\email[]{}

\author[0000-0002-3461-0320]{J.~Silber}
\affiliation{Lawrence Berkeley National Laboratory, 1 Cyclotron Road, Berkeley, CA 94720, USA}
\email[]{}

\author{D.~Sprayberry}
\affiliation{NSF NOIRLab, 950 N. Cherry Ave., Tucson, AZ 85719, USA}
\email[]{}

\author[0000-0003-1704-0781]{G.~Tarl\'{e}}
\affiliation{University of Michigan, 500 S. State Street, Ann Arbor, MI 48109, USA}
\email[]{}

\author{B.~A.~Weaver}
\affiliation{NSF NOIRLab, 950 N. Cherry Ave., Tucson, AZ 85719, USA}
\email[]{}

\author[0000-0002-6684-3997]{H.~Zou}
\affiliation{National Astronomical Observatories, Chinese Academy of Sciences, A20 Datun Road, Chaoyang District, Beijing, 100101, P.~R.~China}
\email[]{}


\begin{abstract}

Using the first spectroscopic data release from the Dark Energy Spectroscopic Instrument (DESI DR1), we conduct the largest systematic search for active galactic nuclei (AGN) signatures in $\approx$1.7 million low-redshift ($0.001 \le z \le 0.45$) line-emitting galaxies. Based on the \nii/$\ha$ versus \oiii/$\hb$ emission-line ratio diagnostic, we identify AGN in 314,245/1,211,573 (25.9\%) high-mass ($\logmass > 9.5$) and 9648/467,214 (2.1\%) dwarf ($\logmass \le 9.5$) galaxies. The AGN fraction in line-emitting galaxies increases monotonically with stellar mass, from $\sim$1.4\% at the low-mass end to $\sim$93.3\% at the high-mass end. Among these AGN, 17,949 exhibit broad $\ha$ emission (BL-AGN), which we use to estimate their black hole (BH) masses via single-epoch virial methods. Leveraging this large BL-AGN sample, we extend the $\mbh - \mstar$ scaling relation down to $\logmass \approx 7.8$ and $\logmbh \approx 4.4$. We find a bimodal distribution of galaxies in this space, characterized by two distinct empirical relations. The relation for galaxies with high stellar and BH masses, extrapolates to overlap with the high-redshift overmassive BH candidates. Our results suggest at least two broad evolutionary pathways for these overmassive BHs and their host galaxies across cosmic time. With this paper, we release the {\tt EmFit} value-added catalog, containing emission-line flux and width measurements for $\sim$7.4 million galaxies, the largest catalog with emission-line decomposition into narrow, broad, and outflow components to date. This work significantly expands upon the early DESI results and provides a statistical sample for probing the galaxy$-$BH connection in the low-mass galaxy regime. 

\end{abstract}



\section{ Introduction \label{sec:intro}}

Dwarf galaxies (stellar mass, $\mstar \le 3 \times 10^{9} \msun$) are considered to be galactic building blocks of massive galaxies in the widely accepted $\Lambda$CDM (Cold Dark Matter + Dark Energy) paradigm of structure formation~\citep[][and references therein]{Springel+2006, Sales+2022}. Identifying and studying black holes (BHs) in nearby dwarf galaxies therefore provides a unique window into the formation of the first galaxies and the earliest BHs in the Universe~\citep[][]{Greene+2020, Reines2022}.

The very presence of BHs in dwarf galaxies was debated until a decade ago, but growing observational evidence now demonstrates that at least a subset of dwarf galaxies host actively accreting BHs, i.e., active galactic nuclei (AGN). Multiwavelength diagnostics, including optical emission-line ratio diagrams \citep{Reines+2013, Moran+2014, Mezcua+2020, Molina+2021, Polimera+2022, Salehirad+2022, Siudek+2023, Mezcua+2024a, Pucha+2025}, infrared color-color diagrams \citep{Kaviraj+2019, Lupi+2020, Latimer+2021b}, radio \citep{Mezcua+2019, Reines+2020, Davis+2022} and X-ray observations \citep{Lemons+2015, Pardo+2016, Mezcua+2016, Mezcua+2018, Birchall+2020, Latimer+2021a, Bykov+2024, Sacchi+2024}, and variability techniques \citep{Baldassare+2020a, Burke+2022, Ward+2022} have together highlighted that dwarf galaxies can indeed host central black holes. Such BHs in dwarf galaxies are prime candidates for Intermediate-mass Black Holes (IMBHs; $\mbh \le 10^{6} \msun$) that are believed to be the relics of first BHs formed in the Universe \citep{Mezcua2017, Greene+2020}. Understanding the demographics of BHs in dwarf galaxies is thus essential for constraining the mechanisms of BH seed formation \citep[][]{Volonteri2010}.

Detailed studies of massive galaxies and their central BHs show that the BH masses correlate with various galaxy properties such as the mass or luminosity of the stellar bulge \citep{Ferrarese&Merritt2000, Marconi&Hunt2003, Lauer+2007, Gultekin+2009, McConnell&Ma2013}, the stellar velocity dispersion ($\sigma_{\star}$) in the bulge \citep[][]{Ferrarese&Merritt2000, Gebhardt+2000, Gultekin+2009, McConnell&Ma2013}, and the galaxy stellar mass \citep{Reines&Volonteri2015, Suh+2020, Pucha+2025}. These scaling relations imply that BHs and their host galaxies grow together through a common physical mechanism, such as AGN feedback, mergers, or secular evolution \citep[][S. M. Jewell et al. 2026, in preparation]{Kauffmann&Haehnelt2000, Somerville+2008, Kormendy&Ho2013, Alexander+2025}. Whether these correlations and the underlying physics extend to the low-mass regime remains unclear.

While the $\mbh - \sigma_{\star}$ relation is considered the most robust of these correlations, it remains poorly constrained in the low-mass regime due to a limited number of galaxies with reliable $\sigma_{\star}$ estimates~\citep[][]{Martin-Navarro&Mezcua2018, Baldassare+2020b}. Measuring $\sigma_{\star}$ in low-mass systems is challenging due to the faintness of these targets and limitations with the current instrumentation. As a result, existing samples are small and limited in redshift coverage.

In contrast, stellar mass measurements can be obtained for a large sample of dwarf galaxies, making the $\mbh - \mstar$ scaling relation ideal for extending the galaxy$-$BH connection to lower galaxy masses and to higher redshifts. However, the distribution of dwarf galaxies in this parameter space remains uncertain. \citet{Reines&Volonteri2015} and \citet{Pucha+2025} found that the AGN in nearby dwarf galaxies agree well with the extrapolation of the scaling relation for massive galaxies. In contrast, recent high-redshift ($z \sim 1-8$) observations reveal overmassive BHs in dwarf galaxies that reside $\sim~1-3$ dex above the local relation~\citep[see][and many others]{Mezcua+2023, Mezcua+2024b, Harikane+2023, Ubler+2023, Maiolino+2024, Juodzbalis+2024}. Although these samples are subject to selection biases that favor luminous AGN and therefore the most massive BHs~\citep{Lauer+2007}, the apparent scarcity of similar overmassive BHs in local dwarf galaxies remains puzzling~\citep{Bustamante-Rosell+2021, Bernal+2025}. It raises important questions about how these high-redshift overmassive BHs and their host galaxies evolve over cosmic time. 

Overall, expanding the census of AGN in dwarf galaxies and extending galaxy$-$BH scaling relations to lower masses are critical for constraining BH seed formation and galaxy$-$BH coevolution~\citep{Greene+2020}. Key questions include the demographics and mass distributions of BHs across different galaxy properties, as well as the efficiency of their coupled growth with host galaxies. Addressing these requires a statistically robust sample spanning a broad range of galaxy and BH properties. 

In this paper, we use the first data release from the Dark Energy Spectroscopic Instrument \citep[DESI; ][]{desi_exp, desi_exp2, desi_dr1} survey to identify AGN in galaxies spanning six orders of magnitude in stellar masses, $6 \le \logmass < 12.5$. As the largest extragalactic spectroscopic dataset available to date, DESI DR1 provides a unique opportunity to construct a statistical sample of AGN in dwarf galaxies and to investigate the interconnection between the galaxies and their central BHs, which is the primary focus of this paper. 

The paper is organized as follows. Section~\ref{sec:data} describes the spectroscopic and photometric data, along with the value-added catalogs (VACs), used in this paper. We present our results related to AGN identification in Section~\ref{sec:active_bhs} and the $\mbh - \mstar$ scaling relation in Section~\ref{sec:mbh-mstar-relation}. We discuss our results in the context of BH seeding mechanisms and galaxy$-$BH coevolution in Section~\ref{sec:discussion}, and conclude in Section~\ref{sec:conclusions}. Throughout this paper, we assume the Chabrier initial mass function \citep[IMF;][]{chabrier_imf}, and the \citet{planck_cosmology} cosmology with $H_{0} =\rm~67.4~km~s^{-1}~Mpc^{-1}$ and $\Omega = 0.315$ . All wavelengths are in vacuum wavelengths, and all magnitudes are in the AB system \citep{AB_Mag_System}.
\section{Data \label{sec:data}}
\subsection{Spectroscopy\label{subsec:spectra}}

The Dark Energy Spectroscopic Instrument (DESI) survey is an eight-year cosmological survey designed to obtain optical spectra of nearly 63 million galaxies and quasars across $\approx17,000~\rm deg^{2}$ \citep{desi_exp, desi_exp2, desi_exp1, desi_ops}. The 5000-fiber multi-object spectrograph covers a spectral range of $3600 - 9800$ \AA, with a resolution ranging from 2000 to 5000 \citep{desi_instrument_overview, desi_focal_plane, desi_corrector, desi_fiber_system}. The DESI Early Data Release \citep[EDR;][]{desi_edr1, desi_edr2} includes all survey validation observations acquired prior to May 2021. The first public data release \citep[DESI DR1;][]{desi_dr1} encompasses the first 13 months of the main survey (2021 May 14 $-$ 2022 June 13), along with a reprocessing of the EDR data. With $\sim$28.4 million spectra of $\sim$27.5 million unique objects, DESI DR1 is the largest extragalactic spectroscopic dataset to date.

DESI targets fall into five primary classes: (1) Milky Way Survey \citep[{\tt MWS};][]{mws}, (2) Bright Galaxy Survey \citep[{\tt BGS};][]{bgs1, bgs2, bgs_wise}, (3) Luminous Red Galaxies \citep[{\tt LRG};][]{lrg1, lrg2}, (4) Emission Line Galaxies \citep[{\tt ELG};][]{elg1, elg2}, and (5) Quasars \citep[{\tt QSO};][]{qso1, qso2}. Additionally, DESI utilizes spare fibers to observe secondary targets \citep[{\tt SCND}; see the Appendix of ][]{desi_edr1, Darragh-Ford+2023, Fawcett+2023}, and also conducts time-sensitive observations of Targets of Opportunity (ToOs) triggered by any transient events. Details regarding the target selection are available in \citet{desi_tgts}. 

DESI spectra are reduced using the DESI spectroscopic pipeline \citep{desi_spec_pipeline}, followed by the {\tt Redrock} redshift-fitting pipeline for obtaining the redshifts of all targets \citep[][S. Bailey et al. 2026, in preparation]{redrock_qso, Anand+2024}. For a subset of {\tt QSO} targets, the initial redshift estimates were found to be unreliable. Therefore, the DESI team used a machine-learning algorithm called {\tt QuasarNet} \citep{quasarnet} to identify quasars and to determine their redshifts accurately~\citep[][]{Alexander+2023, Lan+2023}. Throughout this work, we adopt the corrected {\tt QSO} redshifts from {\tt QuasarNet} and the {\tt Redrock} redshifts for all other targets. 

\subsection{Photometry\label{subsec:photometry}}

The DESI primary targets are selected based on the ninth data release of the DESI Legacy Imaging Surveys \citep[LS DR9;][D. Schlegel et al. 2026, in preparation]{desi_imaging}. All the images from the Legacy Surveys are processed using the {\it Tractor \footnote{\url{https://github.com/dstndstn/tractor}}} code \citep{Lang+2016}, which uses inference modeling to optimize the source morphology and to estimate photometry in the $g$, $r$, and $z$ bands. Using this optical model, {\it Tractor} also performs forced photometry on the  WISE/NEOWISE \citep[unWISE; ][]{Lang+2014, Meisner+2016, Meisner+2017} coadd images (in W1, W2, W3, and W4 bands) from the Wide-field Infrared Space Explorer \citep[WISE; ][]{wise}. 

The LS DR9 catalog thus includes optical and mid-infrared photometry, along with extinction values in linear units of Milky Way transmission for all the sources \citep[][]{extinction_sfd98, extinction_fitzpatrick1998}. It also includes morphological classifications ({\tt TYPE}) and S{\'e}rsic index ({\tt SERSIC}) as estimated by {\it Tractor} for each object. In this paper, we define point sources as galaxies with {\tt TYPE == PSF} and late-type sources as galaxies with {\tt TYPE} as either round-exponential ({\tt REX}), an exponential ({\tt EXP}), or a S\'ersic ({\tt SER}) with a S\'ersic index $\le 2$. The remaining galaxies that are neither point sources nor late-type galaxies are labeled as early-type galaxies.

Some secondary targets and ToOs lack corresponding LS DR9 photometry. Because our sample selection relies on photometric and morphological information (see Sections~\ref{subsec:sample} and~\ref{subsec:mbh-mstar}), we focus only on DESI targets with available LS DR9 photometry.  

\subsection{Stellar Masses \label{subsec:mstar}}

We utilize stellar masses from the DESI DR1 physical properties\footnote{\url{https://data.desi.lbl.gov/doc/releases/dr1/vac/cigale/}} VAC \citep{Siudek+2024}, generated for all galaxies with $z > 0.001$ using Code Investigating GALaxy Emission \citep[{\tt CIGALE} v.22.1; ][]{cigale}. \citet{Siudek+2024} performed the spectral energy distribution (SED) fitting for all DESI targets that are classified as {\tt GALAXY} or {\tt QSO} by {\tt Redrock} and have valid LS DR9 photometry (Section~\ref{subsec:photometry}). They adopted \citet{bc03} single stellar population models with a Chabrier IMF~\citep{chabrier_imf}, solar metallicity, a delayed star formation history with an optional exponential burst, standard nebular emission from \citet{Inoue2011}, dust attenuation model using the \citet{calzetti_dust} attenuation curve, dust emission from \citet{Dale+2014}, and the AGN emission templates from \citet{Fritz+2006}.

{\tt CIGALE} fits the galaxy and AGN components simultaneously using all available photometry ($g$, $r$, $z$, W1, W2, W3, and W4) at the spectroscopic redshift of the galaxy. The AGN fraction is allowed to be zero if the fit is consistent with a negligible AGN contribution. Although mid-infrared photometry improves the fit, {\tt CIGALE} still converges to a reliable solution using confident optical photometry (in $grz$) and any available WISE photometry. The final VAC includes stellar masses, star formation rates (SFRs), and the relative contribution of the dusty torus to the total infrared luminosity. These properties and their uncertainties are estimated as the likelihood-weighted mean and standard deviation of the probability distribution function, respectively. Further details are available in \citet{Siudek+2024}.

\subsection{Emission Line Measurements\label{subsec:emlines}}

We utilize emission-line measurements from the {\tt EmFit}\footnote{\url{https://data.desi.lbl.gov/doc/releases/dr1/vac/emfit/}} VAC~\citep[see ][and Appendix~\ref{app:emfit}]{Pucha+2025} for all analyses presented in this paper (Sections~\ref{sec:active_bhs} and \ref{sec:mbh-mstar-relation}). {\tt EmFit} is a Python-based emission-line fitting pipeline designed for low-redshift ($z \leq 0.45$) DESI galaxies. The code models \hb, \oiiilamlam, \niilamlam, \ha, and \siilam~emission lines using one or more Gaussian components constrained to have positive flux. It independently tests for additional components in the narrow \oiii~and \sii~emission lines, which can either be an extra narrow peak or an outflow component  \citep[see Appendix A in ][]{Pucha+2025}. When each \sii~emission line is best-fit with two components (be it dual narrow peaks or narrow$+$outflow components), the \sii~profile is used as a template for fitting \nii, $\ha$, and $\hb$ lines. The code further tests for an extra broad Balmer component in $\ha$, requiring a minimum Full Width Half Maximum (FWHM) $\ge 300~\kms$. The overall $\ha$ profile is then used as a template for fitting the $\hb$ emission line. Additional components are accepted only if they improve the fit with at least a $5\sigma$ confidence. A complete description of the fitting procedure is provided in \citet{Pucha+2025}. 

The {\tt EmFit} VAC contains fluxes and widths of all primary, secondary (second narrow peak or outflow component), and broad Balmer components, along with the associated uncertainties, that are available in a given galaxy spectrum. We summarize the DESI DR1 EmFit VAC and its data model in Appendix~\ref{app_sub:data_model}. We also illustrate the diversity of fits (Appendix~\ref{app_sub:diversity}) and list known issues and cautionary notes (Appendix~\ref{app_sub:issues}).

\subsection{Sample Selection}\label{subsec:sample}

To identify and study AGN signatures in DESI galaxies, we construct a starting sample with reliable redshifts, stellar masses, and emission-line measurements. We combine all the catalogs described earlier (Sections~\ref{subsec:spectra}$-$\ref{subsec:emlines}) and apply the selection and quality cuts that are described below. 

Starting from the $\sim$27.5 million unique targets in DESI DR1, we use the {\tt ZCAT\_PRIMARY} column to select the ``best'' spectrum per object and remove sources with fiber issues and unreliable redshifts\footnote{{\tt COADD\_FIBERSTATUS = 0} and {\tt ZWARN = 0 or 4}}, yielding $\sim$21.8 million objects. We then select those that are classified as {\tt GALAXY} or {\tt QSO} by {\tt Redrock} in the redshift range of 0.001 $\leq z \leq$ 0.45, resulting in a spectroscopic sample of 7,434,906 sources. 

The LS DR9 catalog includes faint sources that are ``shredded'' from larger galaxies, leading to their misidentification as individual low-mass galaxies. To mitigate this, we use the {\tt FRACFLUX} parameter~\citep[][]{Darragh-Ford+2023}, which quantifies the fraction of flux contributed by nearby sources. We apply cuts on {\tt FRACFLUX}, together with signal-to-noise ratio (SNR) cuts in the optical bands, to remove false detections, extremely faint sources, and fragmented objects:

\begin{equation*}
    {\rm SNR \ge 5~in~}g,r,z~{\rm bands}
\end{equation*}
\begin{equation*}
    {\rm \tt FRACFLUX \le 0.25~in~}g,r,z~{\rm bands}
\end{equation*}

This leaves 6,856,046 galaxies with robust photometry and {\tt CIGALE}-derived stellar masses (Section~\ref{subsec:mstar}). We further select sources with confident stellar masses using:

\begin{equation*}
    \chi_{\rm\tt CIGALE}^{2} \le 10
\end{equation*}
\begin{equation*}
    \logmass \ge 6
\end{equation*}
\begin{equation*}
    {\rm Error~in}~\logmass \le 0.5~{\rm dex}
\end{equation*}

Additionally, {\tt CIGALE} provides two stellar-mass estimates: one from the best-fit model ({\tt best}) and another from the posterior probability density function ({\tt bayes}). The comparison between the two is quantified using {\tt FLAG\_MASSPDF} and is defined as $M_{\tt best}/M_{\tt bayes}$. Following the recommendation from the {\tt CIGALE} VAC~\citep[see][]{Mountrichas+2021, Siudek+2024}, we apply the following cut:

\begin{equation*}
    0.2 \le {\tt FLAG\_MASSPDF} \le 5
\end{equation*}

This yields 6,143,599 galaxies with stellar masses in the range of  $6 \le \logmass \le 13.6$, with a median of $\logmass \approx 10.3$. Of these, 995,885 are dwarf galaxies with $\logmass \le 9.5$, which is a commonly accepted threshold in separating dwarf and high-mass galaxies in dwarf AGN studies~\citep{Reines+2013, Polimera+2022, Pucha+2025}. We note that sources with a point-like morphology ({\tt TYPE = PSF}) may be dominated by quasar/AGN emission and can have unreliable stellar masses~\citep[][]{Pucha+2025}. We include these 171,799 point sources in our analysis but consider them separately when studying the $\mbh - \mstar$ scaling relation (Section~\ref{sec:mbh-mstar-relation}).

Among the 6,143,599 galaxies with stellar masses, 6,099,572 have {\tt EmFit} measurements (Section~\ref{subsec:emlines}). The remaining galaxies lack reliable fits due to noisy spectra and were therefore excluded from the VAC (see Appendix~\ref{app:emfit} for details). In some galaxy spectra, the \sii~(and therefore \nii, $\ha$, and $\hb$) and \oiii~emission lines can independently exhibit an additional component (either a second narrow peak or an outflow). For galaxies with double-peaked emission lines (i.e., {\tt SII\_DBL\_FLAG} or {\tt OIII\_DBL\_FLAG} = True), we sum the fluxes of the two components and propagate the uncertainties in quadrature~ \citep[see Appendix A of][and Appendix~\ref{app:emfit}]{Pucha+2025}. For all other galaxies, we consider only the primary component fluxes and their uncertainties for analysis.

\begin{figure}
    \centering
    \includegraphics[width=\columnwidth]{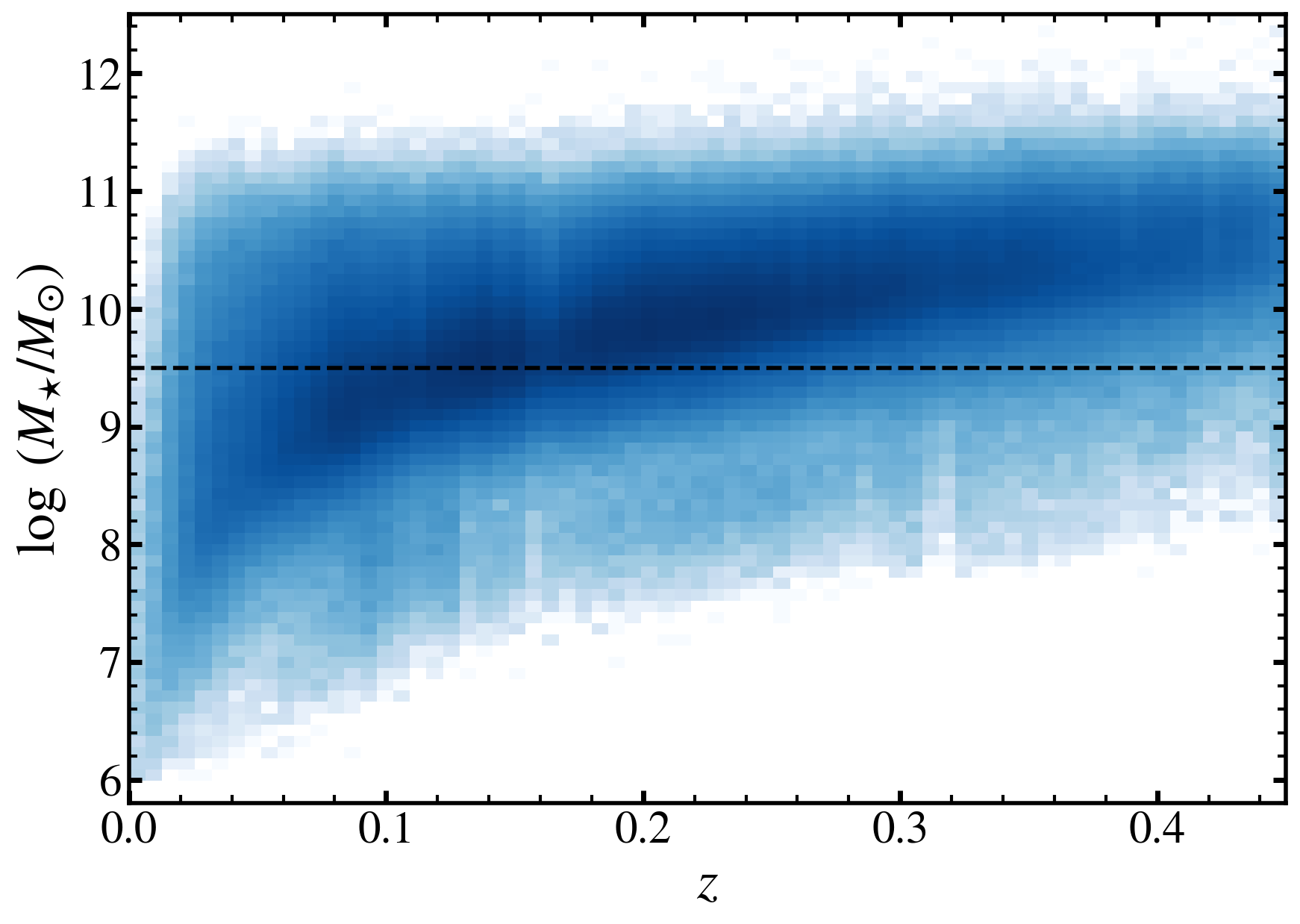}
    \caption{2D density distribution of all 1,678,787 line-emitting galaxies in the $\mstar - z$ space. The horizontal dashed line marks the separation of dwarf galaxies ($\logmass \le 9.5$) and high-mass ($\logmass > 9.5$) galaxies. The galaxies span a broad range of stellar masses ($6 \le \logmass < 12.5$) and redshifts ($0.001 \le z \le 0.45$).} 
    \label{fig:mstar_z_hist}
\end{figure}

We find that 35,666 (0.6\%) galaxies have a second component in \siilam~emission lines, of which 24,735 are double-peaked emission lines. Additionally, 71,618 (1.2\%) galaxies have a second component detected in the \oiiilamlam~lines, of which 15,702 are double-peaked lines. The remaining sources with second components are interpreted as galaxies with outflow signatures and are studied in more detail in an accompanying paper~\citep[][]{Rodriguez+2026}. 

Using the final emission-line fluxes and uncertainties, we select line-emitting galaxies by applying the following criteria on the SNR and amplitude-over-noise (AoN):

\begin{center}
    SNR $\ge$ 3 for \oiii, $\ha$, \nii \\
    (SNR $\ge$ 1) \& (AoN $\ge$ 1) for $\hb$ \\
\end{center}

$\hb$ is often faint in galaxies, and setting a higher threshold can remove good AGN candidates~\citep{CidFernandes+2010}. We therefore adopt a lower SNR threshold for $\hb$ compared to the other emission lines. The additional AoN requirement ensures that the $\hb$ amplitude is reliably measured relative to the rms of the continuum around it. 

Applying these criteria results in 1,678,787 ($\approx$27.5\% of the {\tt EmFit} VAC; see Table~\ref{tab:bpt-numbers}) line-emitting galaxies, including 467,214 dwarf galaxies and 1,211,573 high-mass galaxies. Figure~\ref{fig:mstar_z_hist} shows the $\mstar - z$ distribution of this sample, with the dashed line separating the high-mass and dwarf galaxies. Among the high-mass galaxies, $\approx$96.6\% are {\tt BGS} targets, with {\tt LRG}s comprising the next largest contribution and all other target classes contributing $<$1\%. On the other hand, $\approx$81.9\% of dwarf galaxies are {\tt BGS} targets, followed by {\tt SCND}, {\tt ELG}, and {\tt QSO} contribution, with only $\approx$0.1\% being {\tt LRG} targets.

Finally, 57,007 galaxies from the entire sample of line-emitting galaxies have a non-zero flux in the broad~$\ha$~component. We focus only on candidates with a statistically significant broad component ({\tt PROB\_BROAD} $\ge$ 80\%). For the rest of the paper, we define broad-line (BL) candidates as line-emitting galaxies that satisfy the following cuts:

\begin{center}
    SNR ($\ha$;b) $\ge$ 3 \\
    SNR ($\sigma_{\ha;b}$) $\ge$ 3 \\ 
    AoN ($\ha$;b) $\ge$ 2
\end{center}

\noindent where SNR ($\sigma_{\ha;b}$) is the ratio of the width of the broad $\ha$ component to its uncertainty. This yields 26,588 BL candidates (3,246 dwarf and 23,342 high-mass galaxies), and we classify the remaining 1,652,199 galaxies as narrow-line (NL) candidates. 

\begin{figure*}
    \centering
    \includegraphics[width = \textwidth]{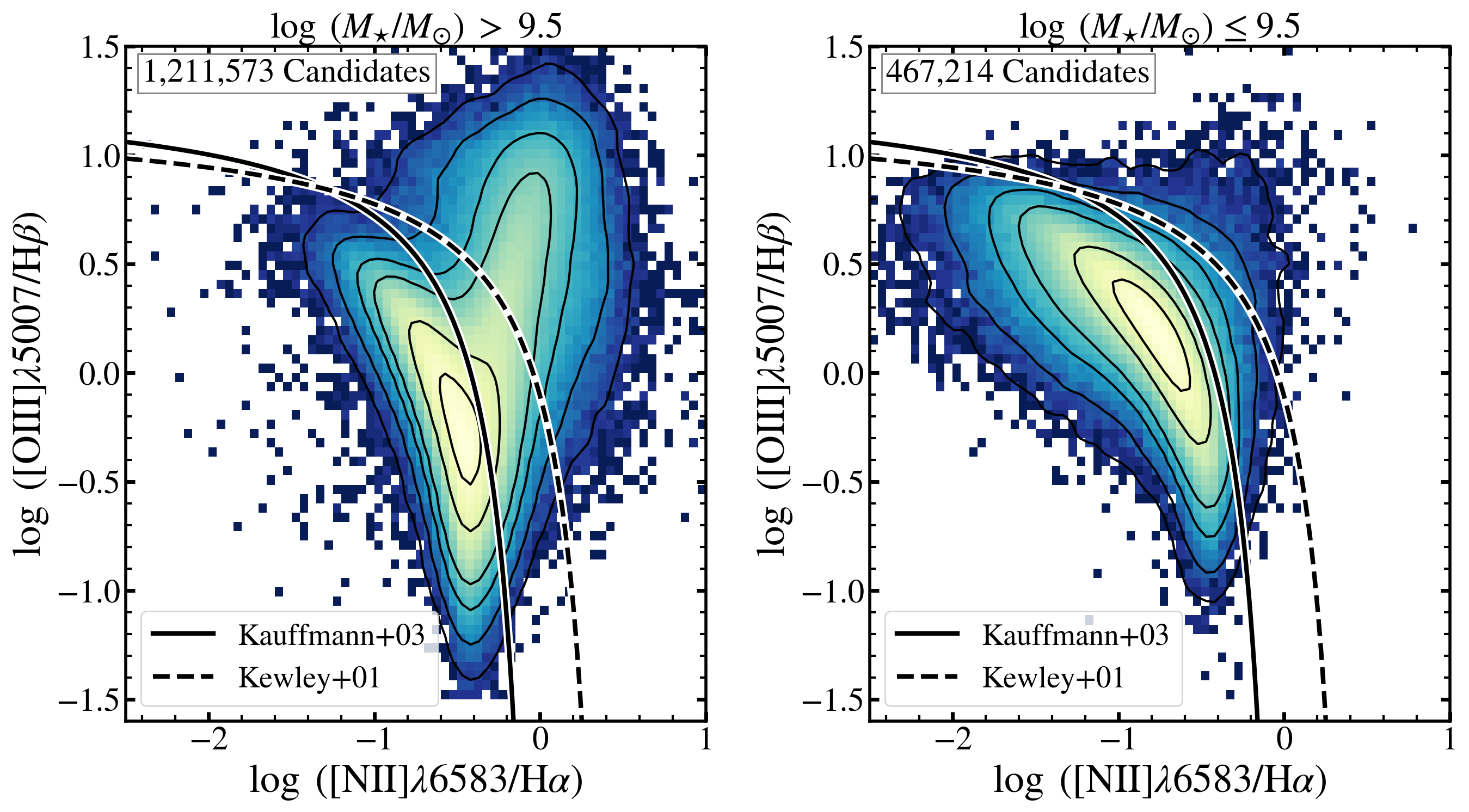}
    \caption{Narrow-line \nii-BPT diagnostic diagram for high-mass ($\logmass > 9.5$; left) and dwarf ($\logmass \leq 9.5$; right) galaxies. Each bin is color-coded by galaxy number density. The bivariate distributions are further shown by the black contours. The thick solid line marks the \citet{Kauffmann+2003} demarcation line, which separates pure star-forming galaxies from those with AGN contribution. The dashed line shows the theoretical ``maximum starburst'' line from \citet{Kewley+2001}. These diagrams reveal 314,245/1,211,573 ($\approx$25.9\%) high-mass AGN and 9,648/467,214 ($\approx$2.1\%) dwarf AGN candidates.}
    \label{fig:bpt_all_2panel}
\end{figure*}

\section{Active Black Holes in Galaxies \label{sec:active_bhs}}

Two-dimensional narrow emission-line ratio diagnostic diagrams, called the Baldwin, Phillips \& Terlevich (BPT) diagrams \citep{bpt, bpt_vo87}, are widely used for separating galaxies with different ionization signatures. These diagnostics separate AGN-dominated galaxies from star-forming galaxies using empirical and theoretical demarcation lines \citep{Kewley+2001, Kauffmann+2003}. 

In this work, we focus on the \nii-BPT diagram with \nii/$\ha$ versus \oiii/$\hb$. This diagnostic is sensitive to gas-phase metallicity, which decreases from the bottom-right corner (high \nii/$\ha$, low \oiii/$\hb$) to the top-left corner (high \oiii/$\hb$, low \nii/$\ha$) of the diagram~\citep[][]{Storchi-Bergmann+1998, Carvalho+2020}. Because dwarf galaxies are typically metal-poor, some dwarf AGN may lie on the star-forming branch of this diagram~\citep{Baldassare+2020a, Molina+2021, Wasleske+2024}. While AGN selected via the \nii-BPT diagram represent an incomplete sample, the diagnostic still remains the most widely used and provides a robust set of candidates across a broad range of stellar masses~\citep{Reines+2013, Reines2022}. Given that our goal is identifying a high-confidence AGN sample rather than a complete one, we restrict our analysis to the \nii-BPT diagnostic diagram in this paper.

\citet{Pucha+2025} reported issues with existing demarcation curves separating AGN from star-forming galaxies in the other two BPT diagrams \citep[i.e., the \sii-~and \oi-BPT diagrams;][]{bpt_vo87}. A more comprehensive analysis using all three diagnostics is underway and will be presented in a future paper. 

In this Section, we quantify the number of AGN and star-forming galaxies in both dwarf and massive galaxies (Section~\ref{subsec:agn-identification}), examine how the AGN fraction varies as a function of stellar mass of galaxies (Section~\ref{subsec:agnfrac}), and discuss its physical interpretation (Section~\ref{subsec:agnfrac_phys}).

\begin{deluxetable*}{lcccc}
\tablecaption{AGN-dominated, Composite, and Star-forming Candidates Identified From NL and BL Candidates. \label{tab:bpt-numbers}}

\tablehead{
\colhead{} & \multicolumn{2}{c}{\bf \underline{Dwarf ($\logmass \le 9.5$) Galaxies}} & \multicolumn{2}{c}{\bf \underline{High-Mass ($\logmass > 9.5$) Galaxies}} \\
All Sources & \multicolumn{2}{c}{987,987} & \multicolumn{2}{c}{5,111,585} \\
Line-Emitting Galaxies & \multicolumn{2}{c}{467,214} & \multicolumn{2}{c}{1,211,573} \\
Non-Line-Emitting Galaxies & \multicolumn{2}{c}{520,773} & \multicolumn{2}{c}{3,900,012} \\
\hline
\colhead{} & \colhead{\bf NL Candidates} & \colhead{\bf BL Candidates} & \colhead{\bf NL Candidates} & \colhead{\bf BL Candidates}
}
\startdata
Line-Emitting Galaxies & 463,968 & 3,246 & 1,188,231 & 23,342 \\
AGN Dominated Sources & 1,455 (0.3\%) & 220 (6.8\%) & 94,036 (7.9\%) & 12,919 (55.3\%) \\
Composites & 7,568 (1.6\%) & 405 (12.5\%) & 201,402 (16.9\%) & 5,888 (25.2\%) \\
Star-Forming & 454,945 (98.1\%) & 2,621 (80.7\%) & 892,793 (75.1\%) & 4,535 (19.4\%) \\
\enddata
\tablecomments{Top: number of line-emitting and non-line-emitting galaxies from the overall parent galaxy sample, separated into dwarf and high-mass galaxies. Bottom: number of AGN-dominated, composites, and star-forming candidates classified using the BPT Diagram of line-emitting galaxies, further divided into NL and BL candidates.}
\end{deluxetable*}

\subsection{AGN Identification}\label{subsec:agn-identification}

Figure~\ref{fig:bpt_all_2panel} presents the \nii-BPT diagrams for high-mass (left panel) and dwarf (right panel) line-emitting galaxies, including both NL and BL candidates (Section~\ref{subsec:sample}). This diagnostic uses only the narrow components of the emission lines. Even for the BL candidates with detected broad Balmer components, the line ratios are estimated using only narrow Balmer components. In both panels, the dashed curve represents the theoretical ``maximum starburst line''  derived using stellar photoionization models \citep{Kewley+2001}. It represents the boundary above which only a non-stellar ionization source (likely an AGN) can explain the observed line ratios. The thick solid line is an empirical line that separates galaxies with AGN contribution from pure star-forming galaxies \citep{Kauffmann+2003}. Sources located between these two curves are classified as ``composites'' due to likely contributions from both star formation and AGN activity. Following previous dwarf AGN studies \citep[e.g.,][]{Reines+2013, Polimera+2022, Salehirad+2022, Mezcua+2020, Mezcua+2024a, Pucha+2025}, we classify both massive and dwarf galaxies in the composite and AGN-dominated regions of the \nii-BPT diagnostic as AGN candidates. 

From Figure~\ref{fig:bpt_all_2panel}, we identify 314,245/1,211,573 (25.9\%) high-mass AGN candidates and 9,648/467,214 (2.1\%) dwarf AGN candidates. Table~\ref{tab:bpt-numbers} details the distribution of AGN-dominated and composite sources among NL and BL sources. We only consider the BL-AGN candidates for the analysis of the \mbh~$-$~\mstar~scaling relation (Section~\ref{sec:mbh-mstar-relation}) but include both NL and BL AGN candidates to compute the fraction of galaxies with an AGN.

The dwarf AGN fraction (2.1\%) is consistent with that observed using DESI early data \citep{Pucha+2025}, still higher than previous estimates based on single-fiber optical spectroscopy~\citep[AGN Fraction $<$ 1\%;][]{Reines+2013, Salehirad+2022}. This is possibly due to DESI's smaller 1.5\arcsec~diameter fiber size compared to the fibers used in these earlier studies (i.e., 3\arcsec~diameter fiber for \citet{Reines+2013} and 2\arcsec~diameter fiber for \citet{Salehirad+2022}), which likely reduces contamination from star-formation~\citep{Pucha+2025}. Similar low dwarf AGN fractions have also been reported in infrared and X-ray studies~\citep[AGN fraction $<$1\%;][]{Mezcua+2018, Birchall+2020, Lupi+2020, Latimer+2021a, Birchall+2022, Bykov+2024}. Despite this higher observed dwarf AGN fraction, the census of AGN in dwarf galaxies remains incomplete. The single-fiber integrated spectroscopic surveys miss AGN associated with wandering and off-nuclear BHs, which can be spatially offset from the galaxy center. Such systems have been identified using integral field unit (IFU) spectroscopy~\citep[][]{Wylezalek+2018, Mezcua+2020, Mezcua+2024a, Bechtold+2026}. 

Nevertheless, the total number of AGN candidates from DESI DR1 data is already four times higher compared to DESI early data \citep{Pucha+2025}, providing strong constraints on the lower limit of the AGN fraction (and thus the BH occupation fraction; see Section~\ref{subsec:agnfrac}). The complete DESI sample, combined with multi-diagnostic AGN selection, will further expand the census of dwarf AGN candidates and will help constrain the galaxy$-$BH connection while accounting for selection effects \citep{Blanton+2026}. 

\begin{figure}
    \centering
    \includegraphics[width = \columnwidth]{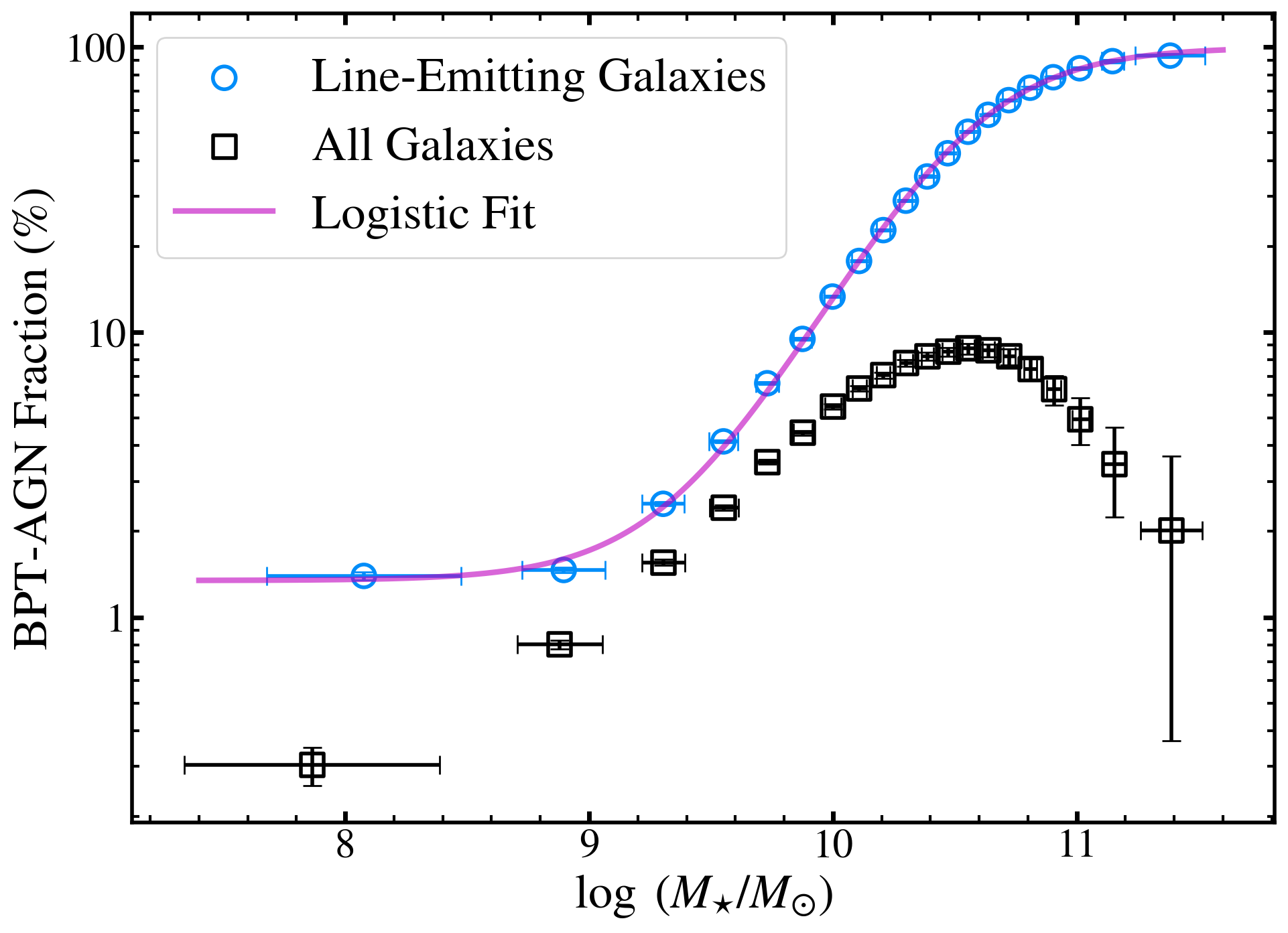}
    \caption{BPT-AGN fraction as a function of stellar mass. The fraction of BPT-AGN among line-emitting galaxies is shown as open blue circles, and the fraction of BPT-AGN candidates in all galaxies is shown as black squares. The logistic fit to the distribution of BPT-AGN fraction among line-emitting galaxies is shown by a pink curve. Overall, the BPT-AGN fraction in line-emitting galaxies increases systematically with stellar mass, while the BPT-AGN fraction when considering all galaxies peaks at $\logmass \approx 10.5$ and declines towards higher masses.}
    \label{fig:agnfrac_mstar}
\end{figure}

\subsection{AGN Fraction\label{subsec:agnfrac}}
The AGN fraction provides a lower limit on the BH occupation fraction, with the low-mass regime offering important insights into the formation of BH seeds in the early Universe \citep[e.g.,][]{Volonteri2010, Mezcua2017, Ricarte+2018, Beckmann+2023}. Light-seed scenarios, i.e., BHs formed as Population~III remnants, predict a high BH occupation fraction in present-day dwarf galaxies, while heavy-seed scenarios, i.e., BHs formed from the direct collapse of gas, predict significantly lower BH occupation values. In this subsection, we construct the lower limit of this BH occupation fraction by examining the fraction of BPT-AGN among galaxies as a function of stellar mass. 

We estimate the BPT-AGN fraction separately for the full galaxy sample (6,099,572 galaxies) and for only line-emitting galaxies (1,678,787 galaxies). We sort all galaxies by stellar mass and divide them into bins of approximately equal number ($\approx$304,980 per bin) of galaxies. Within each bin, we calculate the BPT-AGN fraction as the fraction of galaxies classified as AGN candidates from the \nii-BPT diagram~(Section~\ref{subsec:agn-identification}). The resulting trends are shown in Figure~\ref{fig:agnfrac_mstar}, where open blue circles denote the BPT-AGN fraction among line-emitting galaxies and open black squares represent the BPT-AGN fraction among the full galaxy sample.

For the line-emitting galaxy sample, we find that the BPT-AGN increases with stellar mass in an S-shaped trend, from $\approx$1.4\% at $\logmass \approx 8.1$ to $\approx$93.3\% at $\logmass \approx 11.4$. We model this using a logistic function: 

\begin{equation}
    \begin{split}
        \rm BPT-AGN~Fraction~(\%) = \\
        \rm \frac{L}{1+e^{-k(\log (M_{\star}/M_{\odot}) - x_{0})}} + C
    \end{split}
\end{equation}

\noindent with best-fit parameters $\rm L = 98.65$, $\rm k = 3.59$, $\rm x_{0} = 10.56$, and $\rm C = 1.35$. The resulting fit, shown as a solid pink curve in Figure~\ref{fig:agnfrac_mstar}, asymptotically approaches $\rm L + C = 100\%$ at the high-mass end. Towards the low-mass end, it plateaus at $\approx$1.2\% towards $\logmass \lesssim 8$, revealing a non-zero AGN fraction in the dwarf galaxy regime. 

Because line-emitting galaxies are only a subsample and not necessarily representative of the full galaxy population, we also compute a lower limit to the overall galaxy AGN fraction. Assuming that galaxies failing the emission-line selection criteria (Section~\ref{subsec:sample}) are strictly non-AGN, we calculate the fraction of BPT-AGN in the same mass bins for the full galaxy sample and obtain a different trend. In this case, the BPT-AGN fraction increases from $\approx$0.3\% at $\logmass \approx 7.9$ to peak at $\approx$8.8\% at $\logmass \approx 10.5$, before declining to $\approx$2\% at $\logmass \approx 11.4$. These are consistent with the results from DESI early data~\citep{Pucha+2025}. The divergence between the two AGN fraction estimates is primarily due to the dependence of emission-line selection on stellar masses, with a lower emission-line detection fraction at both the lowest ($\logmass < 8.5$) and highest ($\logmass > 10.5$) stellar masses. We discuss the physical interpretation of these observations, along with the role of selection effects, in the following subsection. 

\subsection{AGN Fraction Physical Interpretation \label{subsec:agnfrac_phys}}

The logistic fit for the observed BPT-AGN fraction (shown as a solid pink curve in Figure~\ref{fig:agnfrac_mstar}) represents the probability of detecting an AGN as a function of stellar mass among line-emitting galaxies. It is a product of several astrophysical and observational factors, as briefly described below:

\noindent{\bf BH Occupation Fraction.} The fundamental requirement for AGN detection is the presence of a central BH. In the hypothetical case where all BHs are actively accreting and detectable by DESI, the observed AGN fraction directly traces the BH occupation fraction. In such a scenario, the declining AGN fraction towards lower galaxy masses could be interpreted as evidence of heavy seeds. However, in reality, the observed AGN fraction provides only a lower limit on the true BH occupation fraction. 

\noindent{\bf AGN Duty Cycle and Accretion.} AGN detection requires the central BHs to be active. This is often quantified in terms of the AGN duty cycle, which is the probability for a given BH to be actively accreting at a given time. Theoretical models suggest that the duty cycle increases with BH mass for $\mbh \ge 10^{6} \msun$ \citep{Shankar+2009, Sabeter+2019, Delvecchio+2020}, although constraints below this mass remain limited. AGN activity depends on gas availability, accretion efficiency, and the poorly-understood Eddington ratio distribution. Recent studies inferring BH occupation fraction using Bayesian modeling found contrasting results in the low-mass regime, primarily due to the difference in the assumptions regarding the underlying accretion rate distributions~\citep{Burke+2025, Zou+2025}. At fixed sensitivity, we can only identify higher Eddington ratio AGN in dwarf galaxies, whereas a broad range of Eddington ratio AGN can be detected in massive galaxies. These factors together likely bias AGN detections towards the most massive BHs and luminous AGN, especially in the low-mass regime. 

\noindent{\bf Emission-Line Selection Effects.} The identification of AGN via emission-line ratio diagnostics needs the efficient detection of emission lines, which has a mass-dependent selection bias. In massive metal-rich galaxies, the \oiii~emission may be weak \citep{Kauffmann+2003}, potentially missing AGN to falling below the \oiii~SNR selection threshold. In contrast, low-mass galaxies tend to be metal-poor and may often fail the \nii~selection criterion \citep{Groves+2006}. Additionally, highly accreting AGN tend to have weaker \oiii~emission due to strong optical \feiilamlam~contamination, further complicating flux measurements~\citep{Panda2024, Wang+2025}. These effects together contribute to the observed divergence between the two BPT-AGN fraction estimates shown in Figure~\ref{fig:agnfrac_mstar}.

\noindent{\bf AGN Diagnostics.} Although the \nii-BPT diagram provides robust AGN identification, it remains incomplete, particularly in dwarf galaxies \citep{Wasleske+2024}. Low-metallicity AGN candidates may occupy the same regions of the \nii-BPT diagnostic as dwarf galaxy starbursts \citep{Groves+2006}. Moreover, studies using unsupervised machine learning techniques showed that the BPT-selected AGN are preferentially found in more massive dwarf galaxies~\citep{Sanjaripour+2025, Wasleske+2025}, indicating that we are likely detecting only the most luminous AGN. 

Ongoing star formation in galaxies can dilute AGN signatures, rendering them undetectable via the BPT diagnostics \citep{Trump+2015, Sanjaripour+2025}. For example, AGN-selected via other diagnostics sometimes fall on the star-forming branch of the \nii-BPT diagram \citep{Baldassare+2020a, Birchall+2020, Molina+2021, Wasleske+2024}.  

In summary, the AGN fraction is shaped by several selection biases, leading to uncertainty regarding the underlying BH population. Nevertheless, both the AGN fraction in line-emitting galaxies and the lower limit on the AGN fraction in all galaxies reveal a non-zero AGN fraction towards the low-mass end, providing a conservative lower limit on the presence of BHs in this regime. It suggests the potential for BH activity beyond the limits of our current stellar-mass selection. 

\begin{figure*}
    \centering
    \includegraphics[width=\textwidth]{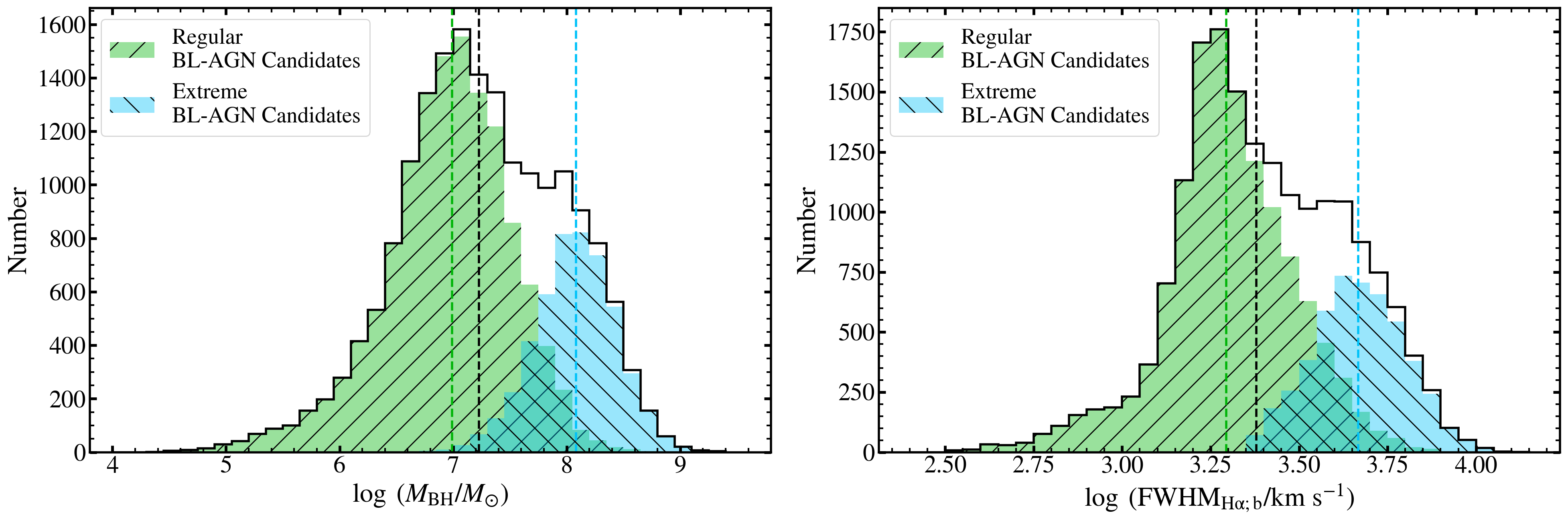}
    \caption{Distribution of BH masses ($\mbh$; left panel) and FWHM of the broad $\ha$ component ($\rm FWHM_{\ha;b}$; right panel) for all 17,949 BL-AGN candidates included in this study (black histograms). Both distributions are bimodal, reflecting contributions from regular and extreme BL-AGN candidates. In both panels, the distributions of regular and extreme BL-AGN candidates are shown by green and blue histograms, respectively. The median values for the full sample are indicated by dashed black vertical lines, while the median values for the regular and extreme BL-AGN sub-samples are marked by dashed green and blue lines, respectively.}
    \label{fig:bhmass_dist}
\end{figure*}

\section{$\mbh - \mstar$ Scaling Relation \label{sec:mbh-mstar-relation}}

Scaling relations between BH masses and host-galaxy properties provide insights into the co-evolution of galaxies and their central BHs. Extending these relations towards low-mass galaxies in the local Universe holds clues related to BH seed formation mechanisms \citep{Volonteri2010, Greene+2020} and BH accretion properties \citep{Ricarte+2018}. Recent observations of overmassive BHs in dwarf galaxies at high-redshifts \citep{Mezcua+2023, Harikane+2023, Ubler+2023, Maiolino+2024, Mezcua+2024b} further raise questions about how such BHs and their hosts grow over cosmic time. In this Section, we present the $\mbh - \mstar$ scaling relation from DESI DR1 and discuss it in the context of these high-redshift observations. 

\subsection{Black Hole Masses \label{subsec:bh_masses}}

The broad components of the Balmer lines provide a way to estimate BH masses using single-epoch virial techniques. These components are thought to originate in the Broad Line Region (BLR) of the AGN. Under the assumption that the BLR gas is virialized, the BH mass can be estimated using the luminosity and width of the broad Balmer component. In this paper, we use the broad \ha~component to estimate the BH masses by adopting the relation derived by \citet{Greene&Ho2005} and modified by \citet{Reines+2013}:

\begin{equation}
    \begin{split}
        \log \left(\frac{\mbh}{\msun}\right) =  \log \epsilon + 6.57 + 0.47\log \left(\frac{L_{\ha;b}}{10^{42}~\rm erg~s^{-1}}\right) \\
        + 2.06\log \left(\rm \frac{FWHM_{\ha;b}}{10^{3}~km~s^{-1}}\right)
    \end{split}
\end{equation}

\noindent where $L_{\ha;b}$ and $\rm FWHM_{\ha;b}$ are the luminosity and FWHM of the broad $\ha$ component and $\epsilon$ is a dimensionless scale factor encapsulating the BLR geometry and kinematics~\citep{Greene&Ho2007, Grier+2013, Grier+2017}. We adopt $\epsilon = 1$ to be consistent with previous studies \citep{Reines&Volonteri2015, Pacucci+2023, Pucha+2025}.

Sources with $\rm FWHM_{\ha;b} < 1000~\kms$ should be considered with caution, as such components may not always trace virialized BLR gas and can instead arise from a missed outflow component or a wrong fit~\citep[see Appendix B of ][and Appendix~\ref{app:emfit} in this paper]{Pucha+2025}. From our parent sample of 19,432 BL-AGN candidates, we select confident BL-AGN candidates based on spectral stacking analysis and visual inspection as described in Appendix~\ref{app:vi}. This yields 17,949 BL-AGN with broad components confidently identified as likely associated with the BLR. Using the broad $\ha$ fluxes and widths from {\tt EmFit}, we estimate BH masses for this sample.

Single-epoch BH mass estimates are reported to have systematic uncertainties of $\sim$0.5 dex, arising from unknown BLR geometry, inclination, kinematics, and the assumed radius-luminosity relation~\citep{McGill+2008, Denney+2009, Park+2012, Shen2013}. We account for this by adding 0.5 dex in quadrature to the propagated errors to compute the final uncertainties in $\mbh$. 

Figure~\ref{fig:bhmass_dist} shows the distributions of BH masses (left panel) and $\rm FWHM_{\ha;b}$ (right panel) for the 17,949 BL-AGN candidates as black histograms. The BH masses of these BL-AGN candidates span a range of $\logmbh = 4.4 - 9.4$, with a median of 7.2. Both distributions exhibit a clear bimodal structure, which we find is associated with the two groups of BL candidates separated by {\tt EmFit}: (1) regular BL-AGN candidates fit with the ``default'' mode and (2) Extreme BL-AGN candidates (EBL-AGN) fit with the ``EBL'' mode \citep[see ][and Appendix~\ref{app:emfit}]{Pucha+2025}. EBL candidates display exceptionally broad $\ha$ components extending to the \sii~region, with median $\rm FWHM_{\ha;b} \approx 4700~\kms$, compared to $\approx 1970~\kms$ for regular BL-AGN candidates.

The distinct distributions could either reflect intrinsic physical differences between the two populations or be an artifact of the fitting procedure (i.e., if the EBL mode systematically overestimates line widths relative to the default mode). To test whether the distinction is caused by the fitting mode, we force {\tt EmFit} to fit EBL-AGN candidates with the default mode and to fit regular BL-AGN candidates with the EBL mode. Recomputing BH masses from these alternative fits, we find that the dual distribution of BL-AGN candidates still persists, indicating that it is likely intrinsic to the populations rather than a consequence of the fitting procedure. 

Of the confident BL-AGN candidates, 13,026 are regular BL-AGN and 4,923 are EBL-AGN. The regular BL-AGN candidates (green histograms in Figure~\ref{fig:bhmass_dist}) span $\logmbh= 4.4 - 8.9$, with a median of 6.9, while the EBL-AGN candidates (blue histograms in Figure~\ref{fig:bhmass_dist}) span $\logmbh = 6.6 - 9.4$, with a median of 8.1. We discuss these dual distributions further in Section~\ref{subsec:mbh_mstar_dual_dist}.

\begin{figure*}
    \centering
    \includegraphics[width=0.95\textwidth]{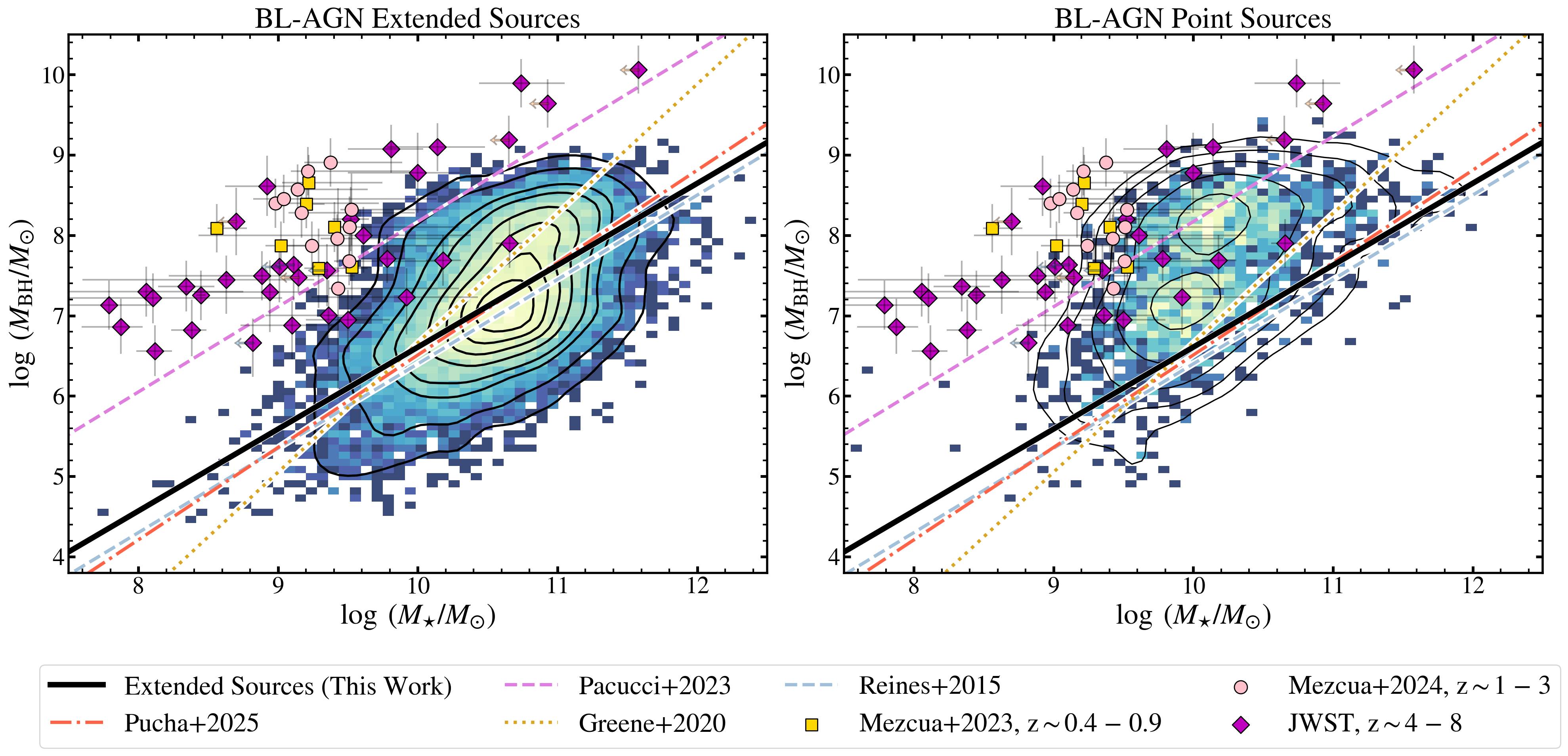}
    \caption{$\mbh - \mstar$ scaling relation of the 17,949 BL-AGN candidates in our sample, separated by morphology. Extended sources ({\tt TYPE} != {\tt PSF}) are shown in the left panel, while point sources ({\tt TYPE} == {\tt PSF}) are shown in the right panel. Logarithmically spaced isodensity contours are overplotted, which highlight the distribution of sources in each panel. Overmassive BHs discovered by \citet{Mezcua+2023} and \citet{Mezcua+2024b} are plotted as yellow squares and light-pink circles, respectively, while magenta diamonds show the high-redshift BHs identified with JWST \citep[][]{Harikane+2023, Kokorev+2023, Larson+2023, Stone+2023, Ubler+2023, Furtak+2024, Juodzbalis+2024, Maiolino+2024, Yue+2024}. In both panels, our empirical fit using only the extended BL-AGN candidates is shown as a solid black line. For comparison, the local scaling relations for active galaxies from \citet{Reines&Volonteri2015} and \citet{Pucha+2025} are plotted as dashed light-blue and dashed-dotted red lines, respectively. The local relation, including inactive galaxies from \citet{Greene+2020}, is shown as a dotted brown line. The high-redshift relation based on a subset of JWST sources from \citet{Pacucci+2023} is plotted as a dashed magenta line. The normalization of the new empirical relation is consistent with early DESI data \citep{Pucha+2025}, while the slope is anchored by the selection and vetting of low-mass BH candidates.} 
    \label{fig:mbh-mstar}
\end{figure*}

At the low-mass end, we identify 792 BL-AGN candidates with $\logmbh < 6$, classified as IMBH candidates. This is more than twice the number identified using early DESI data alone~\citep{Pucha+2025}, which itself already doubled the pre-DESI results. Among these, 50 BL-AGN candidates have $\logmbh < 5$. Combined with the dwarf AGN candidates, this expanded BH population will enable detailed studies of BH demographics and galaxy-BH connection in the low-mass regime. 

\subsection{The Empirical Relation\label{subsec:mbh-mstar}}

Using stellar masses (Section~\ref{subsec:mstar}) and the estimated BH masses (Section~\ref{subsec:bh_masses}) of the 17,949 BL-AGN candidates, we construct the $\mbh - \mstar$ scaling relation, separated by their morphology (Figure~\ref{fig:mbh-mstar}). We treat the extended and point sources separately because the stellar masses of point sources are highly uncertain\footnote{The spectra of point sources are dominated by AGN emission with negligible contribution from the host galaxy, resulting in unreliable stellar mass estimates.}, although their BH masses are robust. As shown by \citet{Pucha+2025}, point sources occupy a distinct region in this space compared to extended sources. We restrict all quantitative analyses to the extended BL-AGN candidates, while showing the point sources separately in the $\mbh - \mstar$ space for completeness.

The left panel of Figure~\ref{fig:mbh-mstar} shows the 15,871 BL-AGN sources with extended morphology ({\tt TYPE} != {\tt PSF}). This includes 276 dwarf AGN candidates, extending the $\mbh - \mstar$ scaling relation down to $\logmass \approx 7.8$ and $\logmbh \approx 4.4$. We fit an empirical relation similar to \citet{Reines&Volonteri2015} and \citet{Pucha+2025} as follows:

\begin{equation}
    \logmbh = \alpha + \beta\log(\mstar/10^{11}\msun)
\end{equation}

\noindent For extended BL-AGN candidates, we find:

\begin{equation}
    \alpha = 7.63~\pm~0.01; \beta = 1.02~\pm~0.01
\end{equation}

\noindent This best-fit relation is shown as a solid black line in both panels of Figure~\ref{fig:mbh-mstar}. For comparison, we overlay the local scaling relations from \citet{Reines&Volonteri2015}, \citet{Greene+2020}, and \citet{Pucha+2025} as dashed light-blue, dotted brown, and dashed-dotted red lines, respectively. 
  
We find a similar normalization as the scaling relation derived from early DESI data \citep[][]{Pucha+2025}, but a higher value relative to \citet{Reines&Volonteri2015}. While all these studies rely on BL-AGN candidates selected via their accretion signatures, the much larger DESI sample, particularly the high density of sources near $\logmass \approx 10.8$ and $\logmbh \approx 7$, provides a more robust constraint on the normalization.

The slope of the relationship is sensitive to the selection and vetting of low-mass BH candidates, which anchor the fit. This is evident from the differences in the final selection of BL-AGN candidates in this regime: \citet{Pucha+2025} considered all statistically significant broad $\ha$ components, including undermassive BH candidates in massive galaxy hosts, while we remove BL-AGN candidates with $\rm FWHM~(\ha;b) < 700~\kms$ for galaxies with $\logmass \gtrsim 10$ based on their stacked spectra lacking a robust broad $\ha$ component (see Appendix~\ref{app:vi}). The slope of our relation deviates at a $3\sigma$ level from \citet{Reines&Volonteri2015}, but $>10\sigma$ from \citet{Pucha+2025}. The reduced uncertainties in our fit reflect the statistical significance of our sample compared to these prior studies. 

\begin{figure*}[t!]
    \centering
    \includegraphics[width=\textwidth]{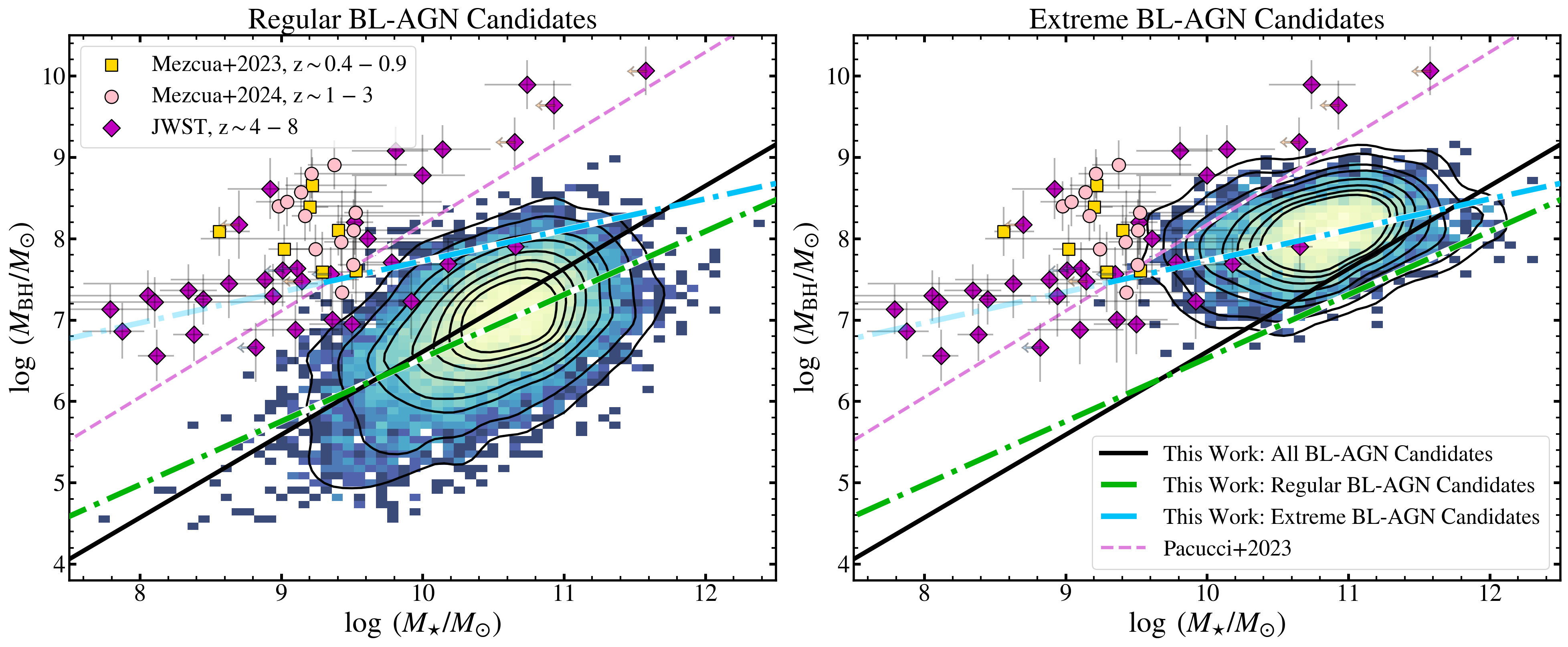}
    \caption{$\mbh - \mstar$ scaling relation of the 11,863 regular extended BL-AGN candidates (left panel) and the 4,008 extreme extended BL-AGN candidates (right panel). The overmassive BHs shown in Figure~\ref{fig:mbh-mstar} are overplotted in both panels. The empirical fit to the full extended BL-AGN sample is shown as a solid black line, while the individual empirical fits for regular and extreme BL-AGN candidates are shown as dashed-dotted green and blue lines, respectively. The high-redshift empirical relation from \citet{Pacucci+2023} is plotted as a dashed magenta line. Notably, the high-redshift BH candidates observed with JWST lie on the low-mass extrapolation of the EBL-AGN relation.}
    \label{fig:mbh-mstar-reg-ebl}
\end{figure*}

These relations also differ from the local relation of \citet{Greene+2020}, which uses a combination of both active galaxies (with single-epoch $\mbh$ estimates) and inactive galaxies (with dynamical BH mass estimates). Because dynamical mass measurements preferentially select the most massive BHs, this sample yields a steeper slope than relations based only on active galaxies \citep{Reines&Volonteri2015}. These differences highlight the need to account for selection effects when interpreting the results.

We further compare our results with observations of low-mass galaxies at intermediate to high redshifts. Recent observations with the James Webb Space Telescope (JWST) have uncovered tens of AGN candidates in low-mass galaxies ($\logmass \sim 8 - 10$) at $z = 4-10$ \citep[e.g.,][]{Harikane+2023, Kokorev+2023, Larson+2023, Stone+2023, Ubler+2023, Furtak+2024, Juodzbalis+2024, Maiolino+2024, Yue+2024}. These candidates have inferred BH masses of $\mbh \sim 10^{7-9} \msun$, placing them one to three orders of magnitude above the expectations from local scaling relations. Some of them are associated with compact, red sources known as Little Red Dots~\citep[LRDs;][]{Maiolino+2024, Yue+2024, Kocevski+2025}, whose AGN nature is debated.  These high-redshift ``overmassive'' BH candidates are shown as magenta diamonds in Figure~\ref {fig:mbh-mstar}. 

Using a subset of these sources \citep[from ][]{Harikane+2023, Maiolino+2024, Ubler+2023}, \citet{Pacucci+2023} derived a high-$z$ empirical relation that deviates from the local relation at a $>3\sigma$ confidence. This relation, shown as a dashed magenta line in Figures~\ref{fig:mbh-mstar}, traces the upper edge of our BL-AGN sample. Recent studies involving statistical modeling of selection effects and measurement uncertainties infer an $\mbh-\mstar$ relation, which is $\sim$1 dex above the local relation, but with a larger intrinsic scatter~\citep{Li+2025, Ziparo+2026}. These studies argue that the underlying relation may remain broadly consistent, while the scatter increases towards higher redshifts.

A similar population of overmassive BHs is also identified at intermediate redshifts. \citet{Mezcua+2023, Mezcua+2024b} report the finding of 19 AGN in low-mass galaxies at $z = 0.35 - 2.7$ with similar overmassive BHs. These sources are overplotted in Figure~\ref{fig:mbh-mstar} as yellow squares and light-pink circles, respectively. We will discuss the case of overmassive BHs further in Section~\ref{subsec:overmassive_dwarf_bhs}.

\subsection{Dual Distribution of BL-AGN Candidates \label{subsec:mbh_mstar_dual_dist}}

The left panel of Figure~\ref{fig:mbh-mstar} shows that the extended BL-AGN candidates exhibit an asymmetric distribution in the $\mbh - \mstar$ space, with an extension towards higher galaxy and BH masses. This asymmetry is even more pronounced in the point source BL-AGN candidates (right panel of Figure~\ref{fig:mbh-mstar}). Previous studies have found that early-type galaxies may have a higher normalization in this space than late-type galaxies~\citep{Reines&Volonteri2015, Greene+2020}. However, we find no such difference in the distributions of early and late-type galaxies in our sample.

We instead find that this asymmetry arises from two subpopulations of BL-AGN candidates, distinguished by their emission-line profiles. Specifically, they are separated by the {\tt EmFit} fitting modes, i.e., the regular BL-AGN and EBL candidates. These two sub-samples are found to result in a bimodal BH mass distribution (see Section~\ref{subsec:bh_masses}), and they occupy distinct locations in the $\mbh - \mstar$ space, as illustrated in the two panels of Figure~\ref{fig:mbh-mstar-reg-ebl}. 

EBL candidates lie systematically between the local scaling relation (solid black line) and the high-redshift relation (dashed magenta line). Similar massive galaxies hosting BHs offset by $\sim$1 dex above the local relation have been reported in both active galaxies via accretion-based searches and inactive galaxies via dynamical searches \citep{Ferre-Mateu+2015, Trakhtenbrot+2015, Walsh+2015, Buchner+2026}. In fact, \citet{Reines&Volonteri2015} showed that inactive galaxies define a relation that lies above that of active galaxies, while \citet{Greene+2020} found a steeper relation when combining a sample of active and inactive galaxies. Some inactive galaxies extend even beyond the EBL-AGN distribution towards higher BH masses. 

We separately fit two empirical relations for regular BL-AGN and EBL-AGN candidates with the same parametrization as mentioned in Section~\ref{subsec:mbh-mstar}: 

\begin{equation}
    \logmbh = \alpha + \beta\log(\mstar/10^{11}\msun)
\end{equation}

\noindent For regular BL-AGN candidates, we find:

\begin{equation}
    \alpha = 7.31~\pm~0.01; \beta = 0.78~\pm~0.01
\end{equation}

\noindent For EBL-AGN candidates, we find:

\begin{equation}
    \alpha = 8.11~\pm~0.01; \beta = 0.38~\pm~0.02
\end{equation}

These fits are shown as dashed-dotted green and blue lines for the regular BL-AGN and EBL-AGN candidates, respectively, in Figure~\ref{fig:mbh-mstar-reg-ebl}. The regular BL-AGN relation has a similar normalization, but shallower slope than the overall extended BL-AGN fit, whereas the EBL-AGN relation is even flatter and lies $\sim$1 dex above them. Interestingly, the high-redshift JWST BH candidates $-$ which were not included in deriving the fit $-$ follow the low-mass extrapolation of the EBL-AGN relation, while some of them extend further above it. 

\begin{figure}
    \centering
    \includegraphics[width=1.0\linewidth]{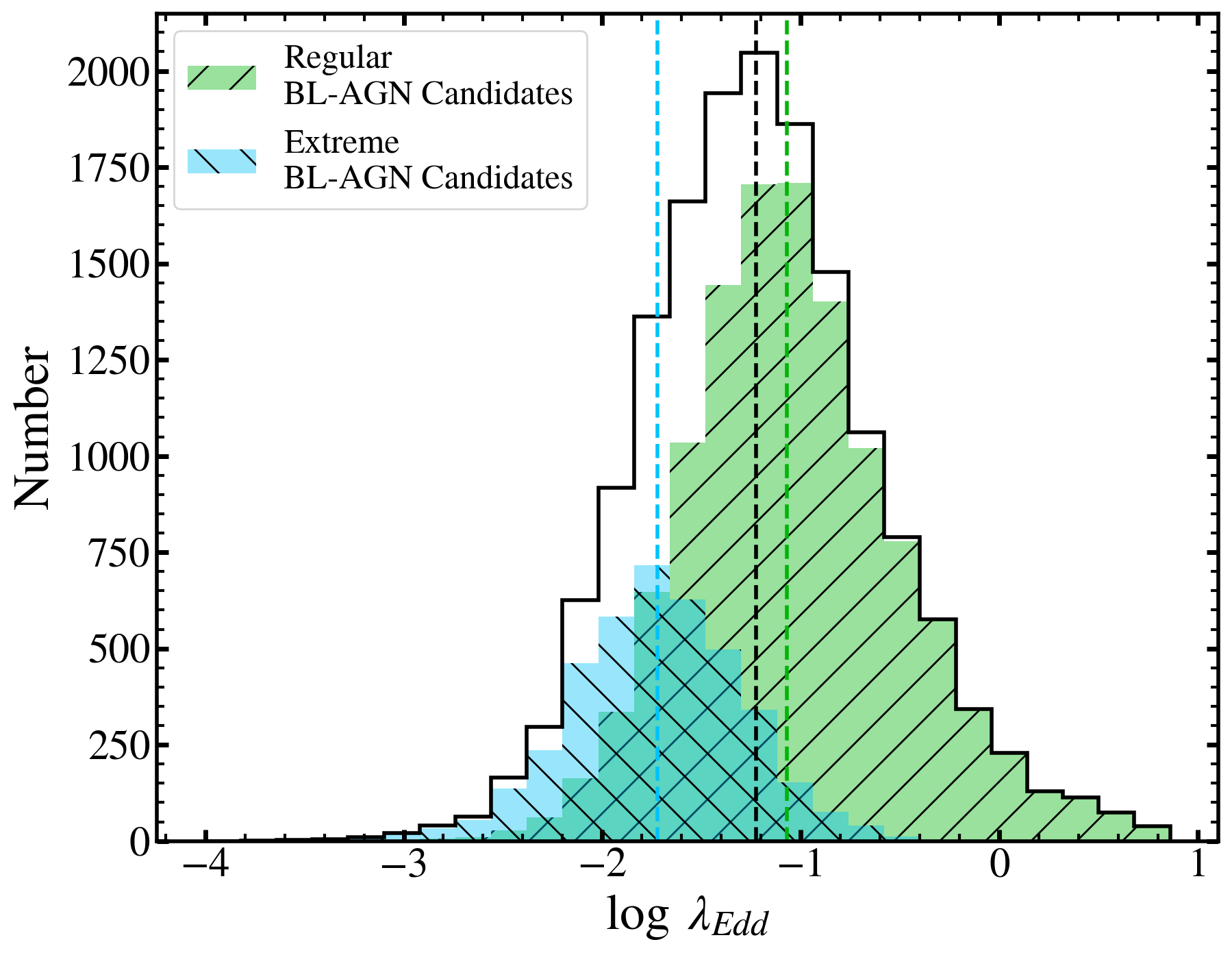}
    \caption{Distribution of Eddington ratios of regular BL-AGN (green histogram) and extreme BL-AGN (blue histogram) candidates. The median values for each population are indicated by corresponding colored dotted vertical lines. On average, the extreme BL-AGN candidates exhibit sub-Eddington ratios and comparatively lower Eddington ratios than regular BL-AGN candidates.}
    \label{fig:lamedd_dist}
\end{figure}

To further investigate the nature of these two subpopulations of BL-AGN candidates, we estimate AGN bolometric luminosities ($L_{\rm bol}$) using \oiii~luminosities \citep[$L_{\rm bol} = L_{\rm \oiii} \times 1000$;][]{Moran+2014}. These range from $\log~(L_{\rm bol}/{\rm ergs~s^{-1}}) \approx 41.1 - 46.4$, with a median of 44.1. Combining these luminosities with their BH masses, we calculate their Eddington ratios ($\lambda_{\rm Edd}$). While \oiii~based estimates can be biased by extinction, star formation, or shocks~\citep{Lamastra+2009}, they enable relative qualitative comparison of $\lambda_{\rm Edd}$ between regular and EBL-AGN sources. As shown in Figure~\ref{fig:lamedd_dist}, EBL-AGN candidates exhibit, on average, lower Eddington accretion ratios (median $\lambda_{\rm Edd}~\approx$ 0.02) compared to regular BL-AGN candidates (median $\lambda_{\rm Edd}~\approx$ 0.08). This is similar to that observed in high-redshift overmassive BH candidates~\citep[][]{Mezcua+2023, Mezcua+2024b, Juodzbalis+2024}. 

We note that these results assume that the broad $\ha$ component in the EBL-AGN candidates originates entirely from the virialized BLR gas. However, {\tt EmFit} models these profiles with a single broad Gaussian (and an extra outflow component based on the \sii~profile in the default mode), which may not adequately capture more complex kinematics (see Figure~\ref{fig:single_gaussian_bl}). In particular, recent near-infrared interferometric observations of a $z \sim 4$ quasar \citep{Gravity+Collab+2026} tracing the spatially resolved kinematics of the $\hb$ emission line in the BLR, show that $>$80\% of the line emission likely originates from outflows. Although such detailed constraints near the BH are not accessible to DESI spectra, they highlight the need for complex modeling of BL profiles to understand BLR kinematics from optical spectra.

\section{Discussion\label{sec:discussion}}

In this section, we discuss the $\mbh - \mstar$ scaling relation in the context of overmassive black holes (Section~\ref{subsec:overmassive_dwarf_bhs}) as well as possible implications for BH seed formation scenarios (Section~\ref{subsec:bh_seeds}) and for galaxy$-$BH co-evolution (Section~\ref{subsec:mbh_mstar_evol}).

\subsection{Overmassive Black Holes in Dwarf Galaxies \label{subsec:overmassive_dwarf_bhs}}

Overmassive BHs, lying well above the local scaling relations, have been observed at intermediate and high-redshifts~\citep{Harikane+2023, Kokorev+2023, Larson+2023, Mezcua+2023, Maiolino+2024, Stone+2023, Ubler+2023, Furtak+2024, Juodzbalis+2024, Yue+2024}. At early epochs, such BHs are attributed to either heavy seed BHs \citep[][]{Scoggins+2023, Bhowmick+2024b, Bhowmick+2025, Natarajan+2024} or to light seed BHs that have undergone multiple periods of super-Eddington accretion \citep[][]{Schneider+2023, Trinca+2026}. Although overmassive BHs are predicted from theoretical simulations \citep[][]{Mezcua+2023, Weller+2023, Zhang+2025}, only a handful have been found at $z < 0.15$ \citep[][]{Bustamante-Rosell+2021, Bernal+2025}. It is currently unclear why overmassive BHs in dwarf galaxies have been observationally elusive in the local Universe. 

In several previous studies, overmassive BHs have been defined as BHs in galaxies with $\mbh/\mstar > 0.01$ \citep[][]{van_Son+2019, Mezcua+2023}. We use the same criterion to quantify this population in our sample. We identify 29/434 ($\sim$6.7\%) dwarf BL-AGN candidates as overmassive BH candidates, when we consider galaxies with both extended and point-like morphologies. However, if we restrict to only extended dwarf galaxies with reliable stellar masses, this number is reduced to only 2/276 ($\sim$0.7\%). Despite the dramatic increase in dwarf AGN candidates from DESI, we still observe a lack of actively accreting overmassive BHs in galaxies with $\logmass \le 9.5$, in contrast with high-redshift observations. Interestingly, the EBL candidates do not meet the standard criterion for overmassive BHs, despite being systematically above the scaling relation for regular BL-AGN candidates. 

However, overmassive BH numbers should be interpreted with caution. The higher overmassive BH fraction among point-like sources likely reflects uncertainty in stellar mass estimates, as AGN emission dominates over the host galaxy. At high redshifts, there are large uncertainties on both the stellar mass and BH mass estimates. For instance, BH masses are typically derived using locally calibrated virial relations, which may not be directly applicable at high redshifts. For LRDs in particular, recent studies suggest that they may represent AGN embedded in a dense gas cloud~\citep[i.e, a BH star;][]{Sun+2026}. The scattered component from the gas can broaden emission lines, increasing the BH mass by $\sim 1 - 2$ orders of magnitude~\citep{Rusakov+2026}. Consequently, some of these high-redshift overmassive BH candidates may be due to systematic biases in the $\mbh$ or $\mstar$ estimates, rather than genuine offsets from the local scaling relation. 

Taken together, our results suggest that overmassive BHs are rare among local low-mass galaxies, at least in extended hosts with more reliable stellar masses. If such objects were to exist locally and actively accreting, DESI should detect at least the most luminous of them. Their apparent scarcity may indicate that a significant fraction of local overmassive BHs are not actively accreting and therefore missed by AGN-based selection, or that the high-redshift candidates evolve toward the local relation by the present day. The latter scenario would result in a decrease in the intrinsic scatter of the $\mbh - \mstar$ relation with cosmic time, as predicted by \citet{Ziparo+2026}. Distinguishing between these possibilities requires detailed follow-up studies of observed overmassive BH candidates across different redshifts to understand their origin and role in galaxy evolution. 

\subsection{BH Seed-formation Mechanisms \label{subsec:bh_seeds}}
The low-mass end of the $\mbh - \mstar$ scaling relation provides an important avenue for distinguishing between different BH seed formation scenarios~\citep{Volonteri+2009, Volonteri2010, Greene+2020}. While these predictions are framed in terms of the $\mbh - \sigma_{\star}$ relation, we can expect a similar qualitative behavior in the $\mbh - \mstar$ space. Light-seed BHs generally predict that the scaling relation established at high masses extends towards the low-mass regime. In contrast, heavy-seed BHs would produce a flattening of the relation at the low-mass end, as more massive seeds will dominate this population. 

The presence of overmassive BH candidates at high redshifts (Section~\ref{subsec:overmassive_dwarf_bhs}) has been interpreted as evidence of either heavy seeds~\citep{Scoggins+2023, Bhowmick+2024b, Bhowmick+2025, Natarajan+2024} or rapid early growth of light seeds via mergers and/or super-Eddington accretion~\citep{Schneider+2023, Bhowmick+2024a, Trinca+2026}. In contrast, our low-redshift sample shows that the best-fit for the $\mbh - \mstar$ scaling relation (Figure~\ref{fig:mbh-mstar}) extends down to $\logmass \approx 7.8$, with no evidence of flattening at the low-mass end. This behavior is consistent with the results from the early DESI data~\citep{Pucha+2025}.

The absence of flattening in the overall relation (Figure~\ref{fig:mbh-mstar}) and the relative scarcity of robust overmassive BH candidates in local dwarf galaxies are broadly consistent with the Population-III seed-formation scenario. However, this interpretation relies on a single underlying scaling relation. If the regular BL-AGN and EBL-AGN candidates trace two distinct scaling relations (Figure~\ref{fig:mbh-mstar-reg-ebl}), it may indicate that both BH seed formation channels occur simultaneously \citep[e.g.][]{Regan+2024}. These interpretations assume that BHs in low-mass galaxies have not undergone significant growth since their formation, a hypothesis that may not hold universally. In fact, \citet{Ricarte+2018} argued that the low-mass end of the scaling relations may be governed by BH accretion modes rather than the initial seed population. Multi-wavelength observations are needed to robustly confirm low-mass BH candidates and characterize their accretion properties, enabling mapping of the low-mass end of the scaling relations and its connection to BH seed-formation models. 

\subsection{Evolutionary Pathways for High-Redshift Overmassive Black Holes \label{subsec:mbh_mstar_evol}}

\begin{figure*}[t!]
    \centering
    \includegraphics[width = 0.9\textwidth]{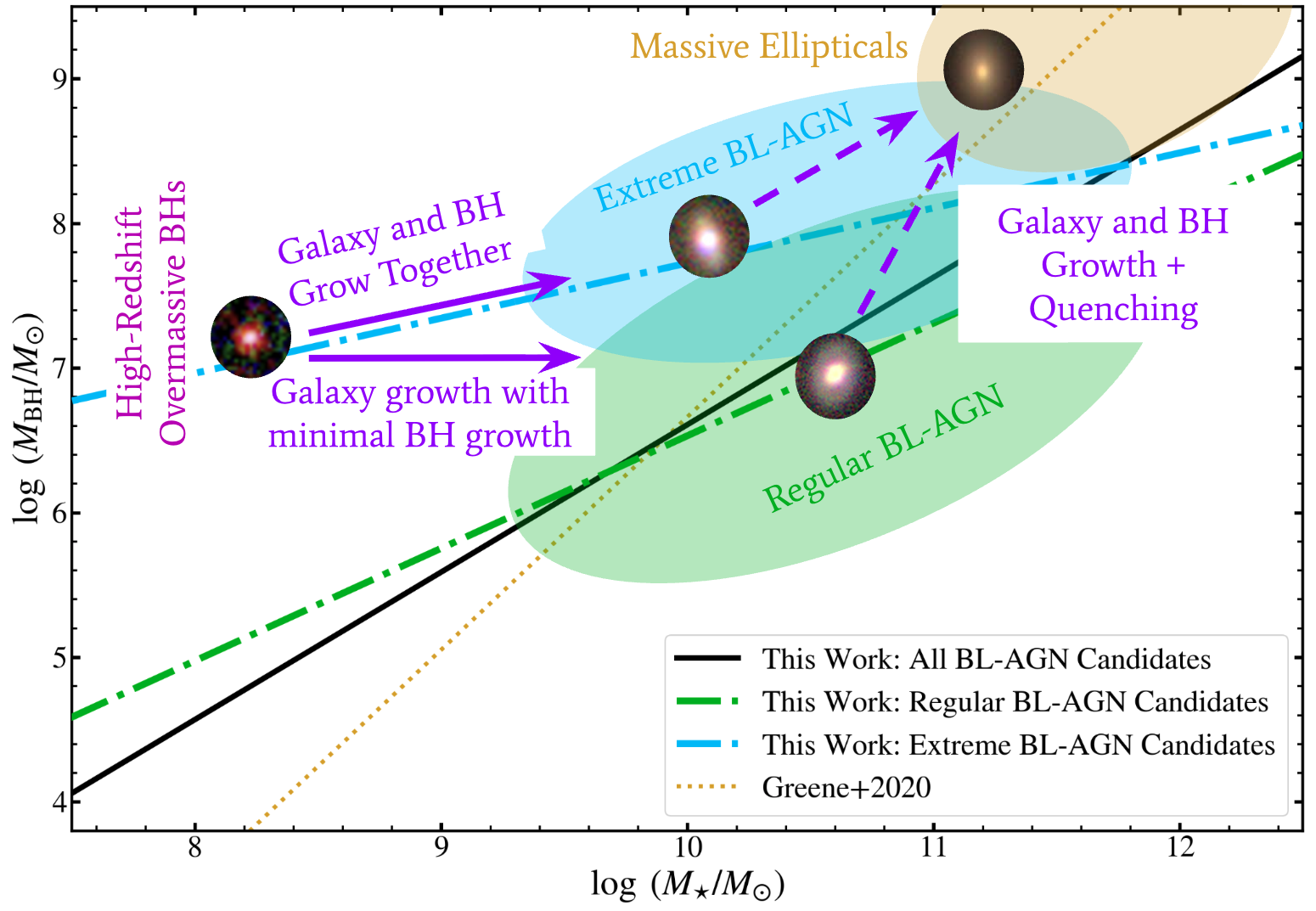}
    \caption{Schematic view of the proposed evolutionary pathways for the high-redshift overmassive BHs and their host galaxies. The host galaxy may grow through star formation and/or mergers with minimal additional BH growth, gradually evolving toward the local $\mbh - \mstar$ relation. Alternatively, the galaxy and the central BH may grow together along the EBL-AGN relation. Some of these systems from either populations may subsequently undergo further galaxy and BH growth, followed by quenching, ultimately forming massive elliptical galaxies hosting massive, inactive BHs.}
    \label{fig:bh-gal-coevol}
\end{figure*}

The observed correlations between BH masses and the host galaxy properties suggest that the galaxies and their central BHs grow together \citep[][]{Kauffmann&Haehnelt2000, Kormendy&Ho2013}. In principle, the tightest correlations are observed between the BH masses and bulge properties such as mass, luminosity, and velocity dispersion \citep[][]{Ferrarese&Merritt2000, Gebhardt+2000, Gultekin+2009, McConnell&Ma2013}. These quantities are difficult to constrain at low galaxy masses and at high redshifts, making the $\mbh - \mstar$ relation more accessible in these regimes. Given that stellar mass reflects a galaxy's overall growth, this space offers a path to study the relative growth of BHs and galaxies. 

As discussed in Section~\ref{subsec:overmassive_dwarf_bhs}, high-redshift surveys have repeatedly identified overmassive BHs lying $\sim 1 - 3$ dex above the local relation \citep{Harikane+2023, Kokorev+2023, Larson+2023, Mezcua+2023, Maiolino+2024, Stone+2023, Ubler+2023, Furtak+2024, Juodzbalis+2024, Yue+2024}. It is possible that these surveys are missing faint, low-mass BHs at these high redshifts due to selection biases~\citep{Lauer+2007}, and are observing only the brightest and most massive BHs. Nevertheless, these observations suggest the existence of at least a population of galaxies at $z \sim 1 - 8$ that host apparent overmassive BHs. 

Overmassive BHs have also been found in massive local galaxies \citep{van_den_Bosch+2012, Emsellem+2013, Ferre-Mateu+2015, Buchner+2026}, and inactive galaxies with dynamical $\mbh$ measurements often lie above the local relation \citep[][]{Reines&Volonteri2015, Kormendy&Ho2013, Greene+2020}. A recent study by \citet{Burke+2024} found that variability-selected AGN at $z~\sim 0.5-4$ occupy the same regions in the $\mbh - \mstar$ plane as inactive galaxies. Our DESI results reveal a dual distribution of the BL-AGN candidates in the $\mbh - \mstar$ space, with a high galaxy and BH mass distribution, lying $\sim$1 dex above the local relation (see Section~\ref{subsec:mbh_mstar_dual_dist}). It is unclear how high-redshift observations of overmassive BHs relate to these massive local galaxies hosting massive BHs. 

\citet{Ferre-Mateu+2015} argue that massive local galaxies with extreme SMBHs are relics of $z \sim 2$ systems that follow a different evolutionary path than the traditional growth of galaxies and BHs. In contrast, \citet{Terrazas+2025} find that present-day galaxies hosting the most massive BHs assembled rapidly at early cosmic times, even before $z \sim 2$. \citet{Stone+2025} studied the future of a single $z \sim 7.08$ quasar by considering the total stellar and gas content in this host and the neighboring galaxies. By summing up these possibilities, they conclude that this galaxy is unlikely to reach the local $\mbh - \mstar$ relation, even without any BH growth. These studies, along with our results, indicate that a single evolutionary track cannot explain the diversity of observed galaxies \citep{Mezcua+2023, Cohn+2025}.

Overmassive BHs may originate in the early Universe, when some BHs undergo rapid, bursty growth, relative to their hosts, while AGN feedback suppresses galaxy growth~\citep{Pacucci+2024, Silk+2024}. Considering such high-redshift overmassive black holes as a starting point, we propose that the two empirical relations traced by regular and EBL-AGN candidates (Figure~\ref{fig:mbh-mstar-reg-ebl}) likely reflect distinct evolutionary paths. 

As schematically illustrated in Figure~\ref{fig:bh-gal-coevol}, the evolution of the high-redshift overmassive BHs may proceed along two broad pathways: (1) Once the bursty growth of the BH is completed, the host galaxy grows through in-situ star formation or dry mergers with minimal BH growth until it ``catches up'' with the BH and reaches the local relation. (2) Alternatively, gas-rich mergers fuel the concurrent growth of the galaxies and BHs, producing EBL-AGN-like systems observed in the local Universe. The diversity of these growth channels likely contributes to the scatter around the observed relations. In some cases, strong AGN feedback during gas-rich phases may quench both star formation and BH accretion, and subsequent dry mergers can lead to the formation of massive elliptical galaxies with massive BHs. 

Semi-analytical simulations of overmassive BHs at $z \sim 1$ by \citet{Mezcua+2023} support this picture. They find that 13\% of the simulated dwarfs with $\logmass \lesssim 9$ remain overmassive at $z \sim 0$ and 35\% evolve into regular massive galaxies ($\logmass > 10$), and the remaining evolve into inactive early-type galaxies. Similarly, using empirical modeling, \citet{Terrazas+2025} predict that nearly all galaxies with $\logmass \gtrsim 10$ that lie on the $\mbh - \mstar$ relation at $z \sim 2$ evolve into quiescent galaxies by $z = 0$.

A comprehensive study of overmassive BHs across cosmic time, coupled with detailed models of AGN feedback and environment effects, is essential for understanding galaxy$-$BH coevolution. This is particularly important in low-mass galaxies, where we can obtain clues regarding the BH seeds. Future DESI data releases, with increased sample size and uniform sky coverage, will enable systematic studies of the impact of AGN-driven outflows and the role of environment on dwarf galaxies and their central BHs.

\section{Conclusions}\label{sec:conclusions}

Using DESI DR1, the largest extragalactic spectroscopic dataset to date, we search for AGN in a sample of 6,143,599 low-redshift ($0.001 \le z \le 0.45$) galaxies spanning a broad range of stellar masses ($6 \le \logmass \le 12.5$). We identify AGN based on optical emission line diagnostics, quantify how the AGN fraction varies with stellar mass, construct the local $\mbh - \mstar$ scaling relation, and study this relation in the context of high-redshift overmassive BHs. Restricting our analysis to the $\sim$1.7 million line-emitting galaxies and separating the sample into dwarf ($\logmass \le 9.5$) and high-mass ($\logmass > 9.5$) galaxies, we conclude the following:

\begin{enumerate}
    \item Based on the optical \nii-BPT emission-line diagnostic, we identify 314,245/1,211,573 (25.9\%) high-mass AGN candidates and 9,648/467,214 (2.1\%) dwarf AGN candidates (Figure~\ref{fig:bpt_all_2panel}). This represents a tenfold increase over the pre-DESI census of optically selected dwarf AGN candidates (Section~\ref{subsec:agn-identification}).

    \item The BPT-AGN fraction, defined as the fraction of AGN-dominated and composite sources among line-emitting galaxies, increases monotonically with stellar mass, rising from $\approx$1.4\% at $\logmass \approx 8.1$ to $\approx$93.3\% at $\logmass \approx 11.4$. This trend is well described by an S-shaped logistic function (Figure~\ref{fig:agnfrac_mstar}), which reflects the AGN detection probability as a function of stellar mass, and depends on several physical and observational factors, including the BH occupation fraction, AGN duty cycle, emission-line selection biases, and the choice of AGN diagnostics. We also report a lower limit on the AGN fraction considering the full galaxy population, which increases from $\approx$0.3\% at $\logmass \approx 7.9$, peaks at $\logmass \approx 10.5$, and then decreases to $\approx$2\% at $\logmass \approx 11.4$ (Section~\ref{subsec:agnfrac}).

    \item Of the 26,588 BL candidates with a broad $\ha$ detection, 19,432 are classified as AGN based on the \nii-BPT diagram (BL-AGN candidates). Among these, 17,949 are candidates with confident broad components likely originating from the BLR (Section~\ref{subsec:agn-identification} and Appendix~\ref{app:vi}). 

    \item We estimate single-epoch virial BH masses for the 17,949 BL-AGN candidates, spanning a range of $\logmbh \approx 4.4 - 9.4$, with a median of 7.2. We find a bimodal distribution of these BH masses, which are attributed to two populations of galaxies that differ based on the width of the broad $\ha$ component: regular BL-AGN (median $\rm FWHM_{\ha;b} \approx 1970~\kms$) and EBL-AGN (median $\rm FWHM_{\ha;b} \approx 4700~\kms$) candidates (Figure~\ref{fig:bhmass_dist}). 
    
    \item We identify 792 IMBH candidates, increasing the census of these elusive objects by more than a factor of two compared to the early DESI data. This expanded sample provides an invaluable dataset for studying BH demographics in the low-mass regime (Section~\ref{subsec:bh_masses}).

    \item Leveraging the large sample of BL-AGN candidates, we extend the $\mbh - \mstar$ scaling relation down to $\logmass \approx 7.8$ and $\logmbh \approx 4.4$ and derive the local empirical relation (Figure~\ref{fig:mbh-mstar}). The normalization is broadly consistent with the prior studies by \citet{Reines&Volonteri2015} and \citet{Pucha+2025}, though the slopes differ. We find that the inferred slope is sensitive to the inclusion of low-mass galaxies and low-mass BHs (Section~\ref{subsec:mbh-mstar}).

    \item We identify a bimodal distribution of BL-AGN candidates in the $\mbh - \mstar$ space. The lower distribution consists of regular BL-AGN candidates, with an empirical fit that has a similar normalization but a shallower slope than the overall scaling relation. The upper distribution is composed of EBL-AGN candidates, with an even flatter relation (Figure~\ref{fig:mbh-mstar-reg-ebl}). On average, EBL-AGN candidates exhibit lower Eddington ratios than regular BL-AGN candidates (Figure~\ref{fig:lamedd_dist}), with predominantly sub-Eddington accretion rates similar to those observed in high-redshift overmassive BHs (Section~\ref{subsec:mbh_mstar_dual_dist}).     

    \item We find a dearth of overmassive BHs ($\mbh/\mstar > 0.01$) in dwarf galaxies in the local Universe. Restricting to dwarf galaxies with reliable stellar masses, only 2/276 ($\sim$0.7\%) host overmassive BH candidates, increasing to 29/434 ($\sim$6.7\%) when point sources are included (Section~\ref{subsec:overmassive_dwarf_bhs}). 

    \item The extension of the $\mbh - \mstar$ scaling relation down to low-mass galaxies without any evidence of flattening, together with the scarcity of overmassive BHs in this regime, is consistent with the Population-III BH seed (i.e., light seeds) formation scenario. However, the two distinct scaling relations for two subpopulations of BL-AGN candidates may indicate that both light seeds and heavy seeds formed simultaneously (Section~\ref{subsec:bh_seeds}).
    
    \item Placing our results in the context of high-redshift overmassive BH candidates, we propose two evolutionary pathways for them: (1) host galaxies grow with minimal BH accretion, moving towards the local $\mbh - \mstar$ relation and the regular BL-AGN distribution, or (2) coevolution of galaxy and the BH leading to the EBL-AGN population. In some cases, subsequent quenching and dry mergers may produce massive elliptical galaxies hosting inactive BHs (Section~\ref{subsec:mbh_mstar_evol} and Figure~\ref{fig:bh-gal-coevol}).

    \item This paper is accompanied by the release of the {\tt EmFit} value-added catalog containing emission-line flux and width measurements of $\sim$7.4 million galaxies (Appendix~\ref{app:emfit}), along with catalogs of identified AGN candidates spanning the full stellar mass range (Figure~\ref{fig:bpt_all_2panel}) and BH masses for the entire BL-AGN sample (Figure~\ref{fig:mbh-mstar}).
\end{enumerate}

The upcoming DESI data releases, combined with multi-diagnostic AGN selection, will enable a comprehensive study of AGN activity across the complete stellar mass range, while accounting for selection effects. Detailed modeling of emission-line kinematics, AGN feedback signatures, and environmental influences will be crucial for interpreting the upcoming results on BH demographics. Ongoing work already includes studies of AGN-driven outflows versus star-formation-driven winds and their roles in shaping galaxy-BH coevolution, especially at the low-mass end of the galaxy mass function. 

\section*{Acknowledgments}

R.P. is currently supported by the University of Utah and was also previously supported by the University of Arizona, and in part by NSF NOIRLab. The research of S.J. and A.D. is supported by the U.S. NSF NOIRLab, which is operated by the Association of Universities for Research in Astronomy (AURA) under a cooperative agreement with the National Science Foundation. M. M. acknowledges support from the Spanish Ministry of Science and Innovation through the projects PID2021-124243NBC22 and PID2024-159201NB-C22. This work was also partly supported by the Spanish program Unidad de Excelencia Mar\'ia de Maeztu CEX2020-001058-M, financed by MCIN/AEI/10.13039/501100011033, and by the MaX-CSIC Excellence Award MaX4-SOMMA-ICE. D.M.A. thanks the Science Technology and Facilities Council (grant codes ST/T000244/1and ST/X001075/1) for support. M.S. acknowledges support by the State Research Agency of the Spanish Ministry of Science and Innovation under the grants 'Galaxy Evolution with Artificial Intelligence' (PGC2018-100852-A-I00) and 'BASALT' (PID2021-126838NB-I00) and the Polish National Agency for Academic Exchange (Bekker grant BPN/BEK/2021/1/00298/DEC/1). This work was partially supported by the European Union's Horizon 2020 Research and Innovation program under the Maria Sklodowska-Curie grant agreement (No. 754510). S.P. is supported by the international Gemini Observatory, a program of NSF NOIRLab, which is managed by the Association of Universities for Research in Astronomy (AURA) under a cooperative agreement with the U.S. National Science Foundation, on behalf of the Gemini partnership of Argentina, Brazil, Canada, Chile, the Republic of Korea, and the United States of America.

This material is based upon work supported by the U.S. Department of Energy (DOE), Office of Science, Office of High-Energy Physics, under Contract No. DE–AC02–05CH11231, and by the National Energy Research Scientific Computing Center, a DOE Office of Science User Facility under the same contract. Additional support for DESI was provided by the U.S. National Science Foundation (NSF), Division of Astronomical Sciences under Contract No. AST-0950945 to the NSF’s National Optical-Infrared Astronomy Research Laboratory; the Science and Technology Facilities Council of the United Kingdom; the Gordon and Betty Moore Foundation; the Heising-Simons Foundation; the French Alternative Energies and Atomic Energy Commission (CEA); the National Council of Humanities, Science and Technology of Mexico (CONAHCYT); the Ministry of Science, Innovation and Universities of Spain (MICIU/AEI/10.13039/501100011033), and by the DESI Member Institutions: \url{https://www.desi.lbl.gov/collaborating-institutions}. Any opinions, findings, and conclusions or recommendations expressed in this material are those of the author(s) and do not necessarily reflect the views of the U. S. National Science Foundation, the U. S. Department of Energy, or any of the listed funding agencies.

The authors are honored to be permitted to conduct scientific research on I'oligam Du'ag (Kitt Peak), a mountain with particular significance to the Tohono O’odham Nation.

The DESI Legacy Imaging Surveys consist of three individual and complementary projects: the Dark Energy Camera Legacy Survey (DECaLS), the Beijing-Arizona Sky Survey (BASS), and the Mayall z-band Legacy Survey (MzLS). DECaLS, BASS and MzLS together include data obtained, respectively, at the Blanco telescope, Cerro Tololo Inter-American Observatory, NSF’s NOIRLab; the Bok telescope, Steward Observatory, University of Arizona; and the Mayall telescope, Kitt Peak National Observatory, NOIRLab. NOIRLab is operated by the Association of Universities for Research in Astronomy (AURA) under a cooperative agreement with the National Science Foundation. Pipeline processing and analyses of the data were supported by NOIRLab and the Lawrence Berkeley National Laboratory (LBNL). Legacy Surveys also uses data products from the Near-Earth Object Wide-field Infrared Survey Explorer (NEOWISE), a project of the Jet Propulsion Laboratory/California Institute of Technology, funded by the National Aeronautics and Space Administration. Legacy Surveys was supported by: the Director, Office of Science, Office of High Energy Physics of the U.S. Department of Energy; the National Energy Research Scientific Computing Center, a DOE Office of Science User Facility; the U.S. National Science Foundation, Division of Astronomical Sciences; the National Astronomical Observatories of China, the Chinese Academy of Sciences and the Chinese National Natural Science Foundation. LBNL is managed by the Regents of the University of California under contract to the U.S. Department of Energy. The complete acknowledgments can be found at \url{https://www.legacysurvey.org/acknowledgment/}.

This research uses services and data provided by the Astro Data Lab, which is part of the Community Science and Data Center (CSDC) program at NSF NOIRLab.

\facilities{Mayall (DESI), Mayall (Mosaic-3), Blanco (DECam), Bok (90Prime), Astro Data Lab, WISE, NEOWISE}

\software{Astropy \citep{astropy+2013, astropy+2018, astropy+2022}, CIGALE \citep{cigale}, FastSpecFit \citep{fastspecfit}, Matplotlib \citep{matplotlib}, NumPy \citep{numpy}, QuasarNet \citep{quasarnet}, Redrock \citep{redrock_qso, Anand+2024}, SPARCL \citep{sparcl}}

\section*{Data Availability}
Supplementary catalogs of identified AGN candidates and the entire BL-AGN sample, along with all the data from the figures, are available in machine-readable form at \url{https://doi.org/10.5281/zenodo.21539891}.


\appendix
\section{EmFit Value-Added Catalog\label{app:emfit}}

We developed a Python-based emission-line fitting code, {\tt EmFit}, to robustly measure emission-line fluxes and widths in low-redshift ($0.001 \le z \le 0.45$) galaxies. The code uses the stellar continuum model from {\tt FastSpecFit}~\citep[][J. Moustakas et al. 2026, in preparation]{fastspecfit}\footnote{\url{https://data.desi.lbl.gov/doc/releases/dr1/vac/fastspecfit/}}, which is subtracted from the DESI spectrum to produce a rest-frame, continuum-subtracted emission-line spectrum. {\tt EmFit} then performs a non-linear least-squares fitting, which focuses on modeling \hblam, \oiiilamlam, \niilamlam, \halam, and \siilam~emission lines. It also tests for additional components in \oiii~and \sii~lines, which can manifest as either double-peaked narrow components or outflow signatures. The dual narrow peaks may be a result of dual AGN, accretion disk kinematics, or complex gas kinematics (e.g., ejecta or bar) within the host galaxy~\citep[][]{Fu+2012, Qiu+2025, Maschmann+2023, Zheng+2025}.

\begin{deluxetable}{cccl}
\tablecaption{EmFit Data Model\label{tab:emfit-datamodel}}
\tablecolumns{4}
\tablewidth{0pt}
\tablehead{
  \colhead{Column Name} & \colhead{Type} & \colhead{Units} & \colhead{Description}
}
\startdata
TARGETID & int64 & -- & Unique identifier of the DESI target \\
SPECPROD & bytes9 & -- & Spectral Production Pipeline ({\tt iron} for DR1) \\
SURVEY & bytes4 & -- & DESI Survey of the spectra: {\tt main}, {\tt special}, {\tt sv1}, {\tt sv2}, {\tt sv3}, or {\tt cmx} \\
PROGRAM & bytes6 & -- & Observing Program: {\tt dark}, {\tt bright}, {\tt backup}, or {\tt other} \\
HEALPIX & int32 & -- & Healpix number of the target \\
Z & float64 & -- & Redshift of the target \\
PROB\_BROAD & float64 & -- & Percentage of broad H$\alpha$ detection in iterations \\
EMLINE\_AMPLITUDE & float64 & $10^{-17}$ \ergsAAcmsq & Amplitude of the emission line component \\
EMLINE\_AMPLITUDE\_ERR & float64 & $10^{-17}$ \ergsAAcmsq & Uncertainty in Amplitude of the emission line component \\
EMLINE\_MEAN & float64 & \AA & Mean of the emission line component \\
EMLINE\_MEAN\_ERR & float64 & \AA & Uncertainty in Mean of the emission line component \\
EMLINE\_STD & float64 & \AA & Standard Deviation of the emission line component \\
EMLINE\_STD\_ERR & float64 & \AA & Uncertainty in Standard Deviation of the emission line component \\
EMLINE\_FLUX & float64 & $10^{-17}$ \ergscmsq & Flux from the emission line component \\
EMLINE\_FLUX\_ERR & float64 & $10^{-17}$ \ergscmsq & Uncertainty in Flux from the emission line component \\
EMLINE\_FLUX\_LERR & float64 & $10^{-17}$ \ergscmsq & 16th percentile Flux from the emission line component \\
EMLINE\_FLUX\_UERR & float64 & $10^{-17}$ \ergscmsq & 84th percentile Flux from the emission line component \\
EMLINE\_SIGMA & float64 & \kms & Width of the emission line component \\
EMLINE\_SIGMA\_ERR & float64 & \kms & Uncertainty in the Width of the emission line component \\
EMLINE\_SIGMA\_LERR & float64 & \kms & 16th percentile width of the emission line component \\
EMLINE\_SIGMA\_UERR & float64 & \kms & 84th percentile width of the emission line component \\
EMLINE\_SIGMA\_FLAG & int64 & -- & Flag regarding instrumental resolution correction for the emission line component \\
WINDOW\_CONTINUUM & float64 & $10^{-17}$ \ergsAAcmsq & Continuum around the emission lines in the fitting window \\
WINDOW\_CONTINUUM\_ERR & float64 & $10^{-17}$ \ergsAAcmsq & Uncertainty in Continuum around the emission lines in the fitting window \\
WINDOW\_NOISE & float64 & $10^{-17}$ \ergsAAcmsq & rms of the Continuum around the emission lines in the fitting window \\
WINDOW\_NDOF & int64 & -- & Number of Degrees of Freedom associated with the fit \\
WINDOW\_RCHI2 & float64 & -- & Reduced chi\textsuperscript{2} associated with the fit \\
OIII\_DBL\_FLAG & bool & -- & \shortstack[l]{Flag regarding whether the \oiii~emission \\ has double narrow-peak Gaussian components} \\
SII\_DBL\_FLAG & bool & -- & \shortstack[l]{Flag regarding whether the \sii~(and therefore \nii, $\ha$, and $\hb$) emission \\ has double narrow-peak Gaussian components} \\
\enddata
\end{deluxetable}

{\tt EmFit} also tests for an extra broad component in the Balmer lines, with minimum $\rm FWHM_{\ha;b} \ge 300~\kms$. It operates in two modes: ``default'' mode and ``Extreme Broad Line (EBL)'' mode, with the latter designed for galaxies where the broad $\ha$ component extends up to the \sii~emission lines. The full fitting procedure for both modes is described in \citet{Pucha+2025}. 

We run {\tt EmFit} on 7,427,711 sources selected after removing objects with fiber and redshift issues (Section~\ref{subsec:sample}). These include DR1 targets classified as {\tt GALAXY} or {\tt QSO} by {\tt Redrock} in the redshift range of  0.001 $\le z \le$ 0.45. After the complete run, we remove sources with poor fits as described in \citet{Pucha+2025}, yielding a final sample of 7,378,347 galaxies. This catalog is publicly available as a VAC\footnote{\url{https://data.desi.lbl.gov/doc/releases/dr1/vac/emfit/}}. 

In this section, we present the VAC data model, illustrate the range of fits, and describe the known issues and caveats of the catalog. 

\begin{table}
    \centering
    \caption{Emission-line components within each fitting window}
    \label{tab:emline_window}
    \begin{tabular}{|c|c|}
    \hline
    {\tt WINDOW}     &    {\tt EMLINE}  \\
    \hline
    {\tt HB}     &  {\tt HB\_N}, {\tt HB\_OUT}, {\tt HB\_OUT} \\
    {\tt OIII}   & {\tt OIII4959}, {\tt OIII4959\_OUT}, {\tt OIII5007}, {\tt OIII5007\_OUT} \\
    {\tt NII\_HA} & {\tt NII6548}, {\tt NII6548\_OUT}, {\tt HA\_N}, {\tt HA\_OUT}, {\tt HA\_B}, {\tt NII6583}, {\tt NII6583\_OUT} \\
    {\tt SII} & {\tt SII6716}, {\tt SII6716\_OUT}, {\tt SII6731}, {\tt SII6731\_OUT} \\
    {\tt HB\_OIII} & {\tt HB\_N}, {\tt HB\_OUT}, {\tt HB\_OUT}, {\tt OIII4959}, {\tt OIII4959\_OUT}, {\tt OIII5007}, {\tt OIII5007\_OUT} \\
    {\tt NII\_HA\_SII} & \shortstack[c]{{\tt NII6548}, {\tt NII6548\_OUT}, {\tt HA\_N}, {\tt HA\_OUT}, {\tt HA\_B}, {\tt NII6583}, {\tt NII6583\_OUT},\\ {\tt SII6716}, {\tt SII6716\_OUT}, {\tt SII6731}, {\tt SII6731\_OUT}} \\
    \hline
    \end{tabular}
\end{table}

\subsection{Data Model \label{app_sub:data_model}}

The full data model for the {\tt EmFit} VAC is available at \url{https://data.desi.lbl.gov/doc/releases/dr1/vac/emfit/#data-model}, and is summarized in Table~\ref{tab:emfit-datamodel}. We outline the key columns below:

\begin{itemize}
    \item The {\tt TARGETID} represents the unique identifier of the DESI target. The columns {\tt SPECPROD}, {\tt SURVEY}, {\tt PROGRAM}, and {\tt HEALPIX} contain information related to DESI targeting and are required to access the spectra of the object. Details regarding these columns are available in \citet{desi_dr1}. The DESI redshifts from {\tt Redrock} after the corrections by {\tt QuasarNet} are listed in the {\tt Z} column.
    
    \item The {\tt PROB\_BROAD} column gives the percentage of iterations in which a broad $\ha$ component is detected.

    \item {\tt EmFit} divides each rest-frame spectrum into fitting windows labeled as {\tt WINDOW} in Table~\ref{tab:emfit-datamodel}. For ``default'' mode, {\tt WINDOW} includes {\tt HB}, {\tt OIII}, {\tt NII\_HA}, and {\tt SII}. For ``EBL'' mode, {\tt WINDOW} includes {\tt HB\_OIII} and {\tt NII\_HA\_SII}.

    \item Within each {\tt WINDOW}, we have multiple components labeled as {\tt EMLINE} in Table~\ref{tab:emfit-datamodel}. The mapping between the {\tt WINDOW} and {\tt EMLINE} is listed in Table~\ref{tab:emline_window}. Secondary narrow or outflow components are labeled by {\tt *\_OUT*} columns, and broad Balmer components are labeled as {\tt *\_B*} columns.
    
    \item For each modeled component, the columns {\tt *\_AMPLITUDE}, {\tt *\_MEAN}, and {\tt *\_STD}, along with their associated uncertainties, record the best-fit Gaussian amplitude, mean, and standard deviation, respectively. The {\tt *\_FLUX} column corresponds to the integrated line flux, estimated as the area under the Gaussian profile.

    \item The observed standard deviation is converted to velocity space and corrected for DESI's instrumental resolution (estimated as the median resolution element in each {\tt WINDOW}) $-$ these values are stored in {\tt *\_SIGMA} columns. The corresponding {\tt *\_SIGMA\_FLAG} indicates:

    \begin{itemize}
        \item[$\bullet$] 0: the component is resolved and corrected for instrumental resolution.
        \item[$\bullet$] 1: the component is unresolved and the reported {\tt *\_SIGMA} is estimated from the best-fit.
        \item[$\bullet$] -1: the component is not detected.
    \end{itemize}
    
    \item If a component is not detected, all the associated columns are set to zero and the {\tt *\_SIGMA\_FLAG} is set to -1.

    \item The {\tt *\_CONTINUUM} and {\tt *\_CONTINUUM\_ERR} provide any additional continuum level within each fitting window.

    \item The {\tt *\_NOISE} is the root-mean-square (rms) of the continuum around the emission lines in the fitting window.
    
    \item The number of degrees of freedom and reduced $\chi^{2}$ for each fitting window are reported in {\tt *\_NDOF} and {\tt *\_RCHI2} columns, respectively.

    \item The {\tt *\_DBL\_FLAG} are boolean flags which indicate whether the corresponding emission line exhibits double narrow-peaks. When this flag is set to True, we recommend adding the fluxes of the two components (i.e., {\tt *\_FLUX} and {\tt *\_OUT\_FLUX}) before using the line flux for any analysis. 
    
    \item The fitting mode (``default'' versus. ``EBL'') can be inferred from the {\tt WINDOW} used. For example, in the ``default'' mode, the EBL-only {\tt WINDOW} columns i.e, {\tt HB\_OIII\_NDOF}, {\tt HB\_OIII\_RCHI2}, {\tt NII\_HA\_SII\_NDOF}, and {\tt NII\_HA\_SII\_RCHI2} are set to zero.  
\end{itemize}

\begin{figure*}[t!]
    \centering

    \begin{subfigure}{1.0\textwidth}
        \centering
        \includegraphics[width=\textwidth]{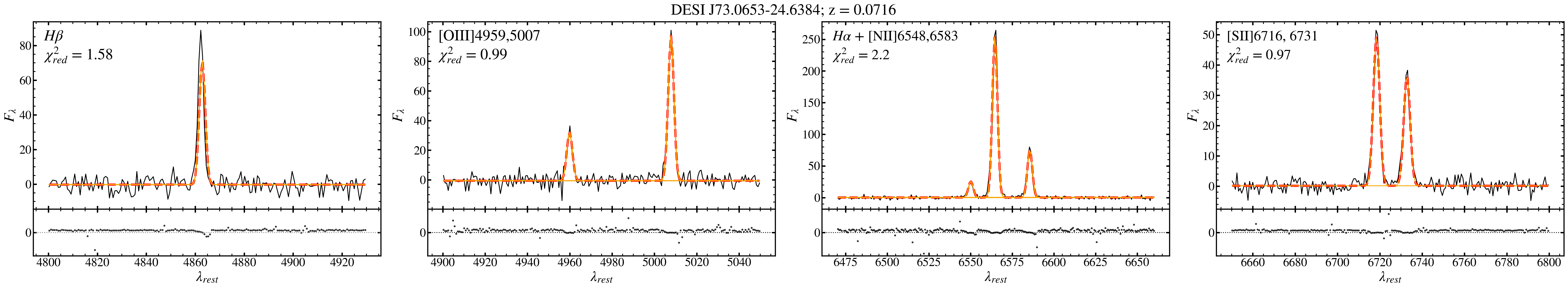}
    \end{subfigure}
    
    \begin{subfigure}{1.0\textwidth}
        \centering
        \includegraphics[width=\textwidth]{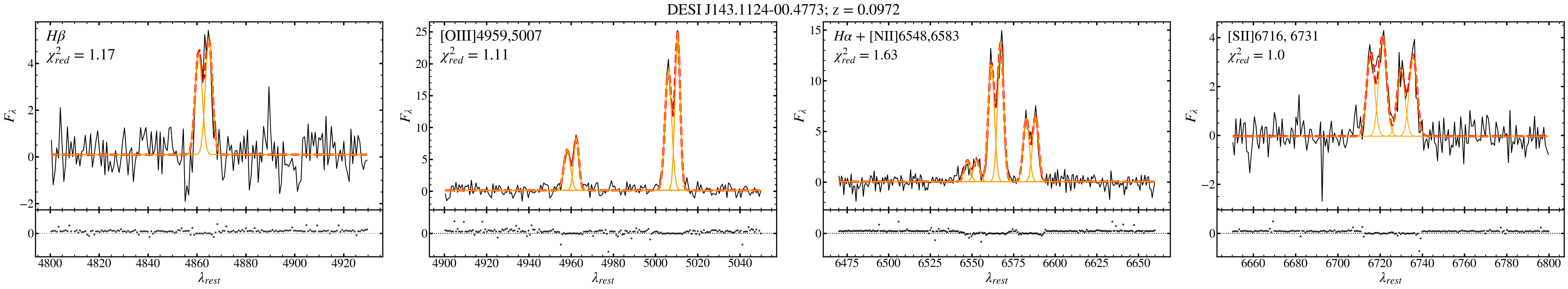}
    \end{subfigure}

    \begin{subfigure}{1.0\textwidth}
        \centering
        \includegraphics[width=\textwidth]{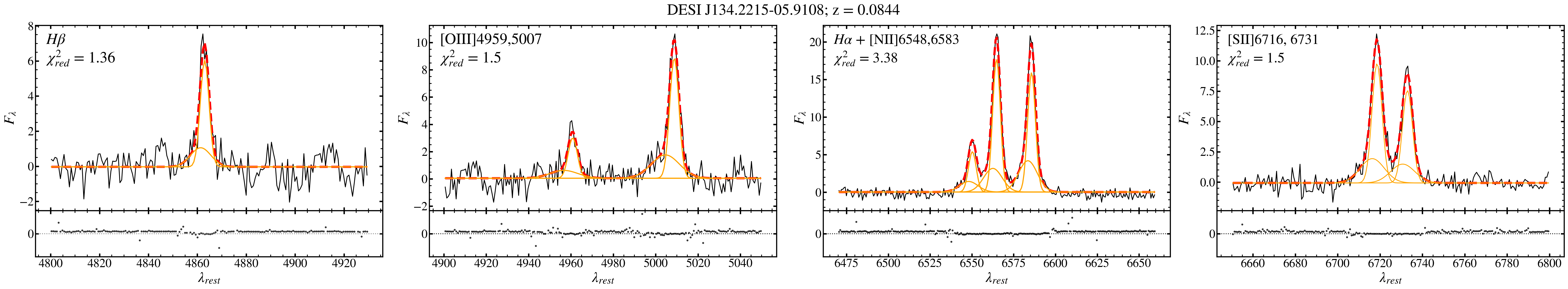}
    \end{subfigure}

\caption{{\bf Example fits of galaxy spectra fit via the default mode without a broad Balmer component.} These include a narrow emission-line only spectrum (top panel), a double narrow-peaked galaxy spectrum (middle panel), and a spectrum with outflow components (bottom panel).  In all the panels, the best-fit models to the continuum-subtracted emission-line spectrum in the regions of $\hb$, \oiii, \nii~+~$\ha$, and \sii~are shown from left to right. The spectra are shown in black, while the best-fit models are shown in dashed red. The individual narrow and extra components are plotted in orange. The reduced $\chi^{2}$ values for each fit are given in the upper-left corner of the individual panels. The fractional residuals are plotted as gray points in the bottom panels for the fits.}
    \label{fig:default_nobroad}
\end{figure*}

\subsection{Diversity of spectra and fits \label{app_sub:diversity}}

\begin{figure*}[h!]
    \centering

    \begin{subfigure}{1.0\textwidth}
        \centering
        \includegraphics[width=\textwidth]{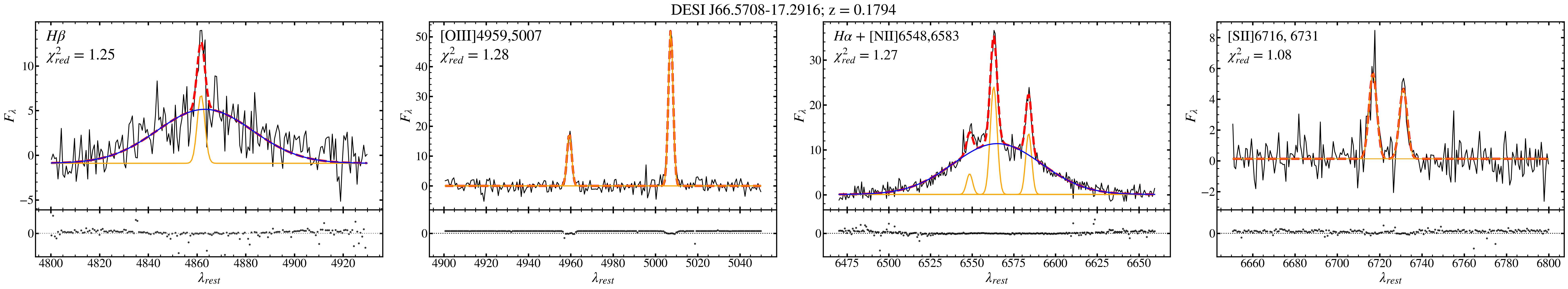}
    \end{subfigure}
    
    \begin{subfigure}{1.0\textwidth}
        \centering
        \includegraphics[width=\textwidth]{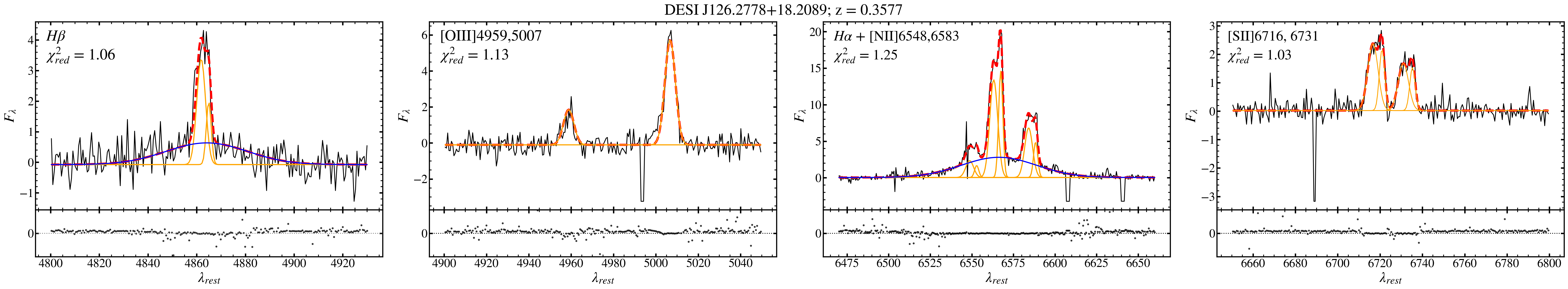}
    \end{subfigure}

    \begin{subfigure}{1.0\textwidth}
        \centering
        \includegraphics[width=\textwidth]{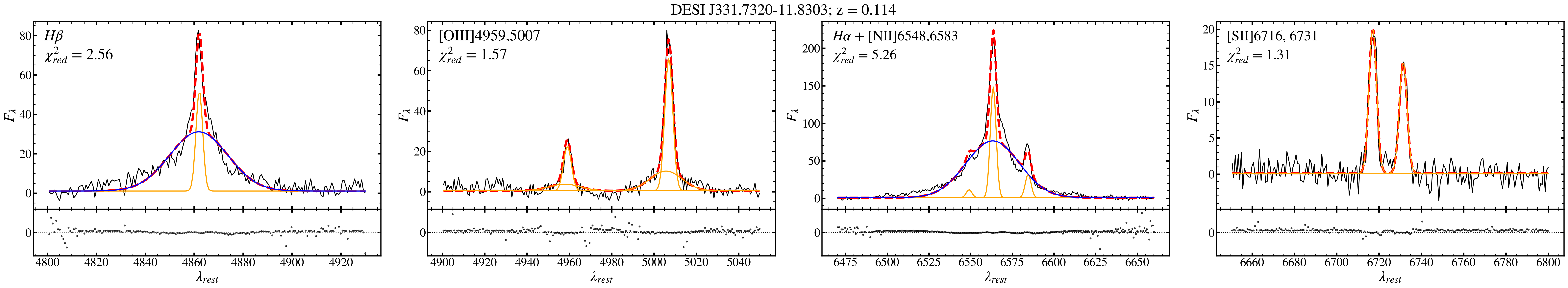}
    \end{subfigure}

\caption{{\bf Example fits of galaxy spectra fit via the default mode with broad Balmer components}, which are shown as blue curves. These include spectra of narrow emission lines only (top panel), double emission lines in all emission lines except \oiii~(middle panels), and an outflow component in only \oiii~line (bottom panel). The colors and text in the panels are the same as Figure~\ref{fig:default_nobroad}.}
    \label{fig:default_broad}
\end{figure*}

\begin{figure*}[h!]
    \centering

    \begin{subfigure}{1.0\textwidth}
        \centering
        \includegraphics[width=\textwidth]{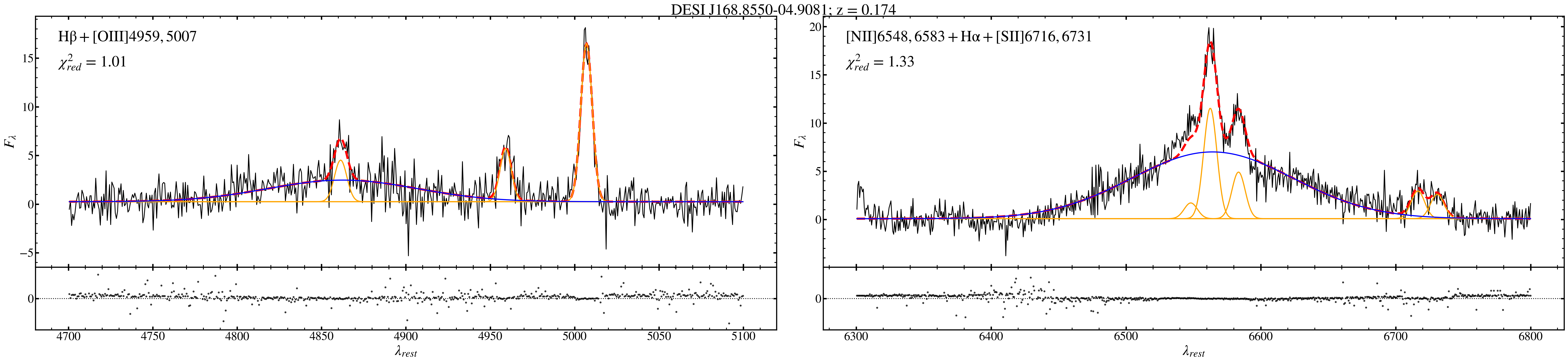}
    \end{subfigure}
    
    \begin{subfigure}{1.0\textwidth}
        \centering
         \includegraphics[width=\textwidth]{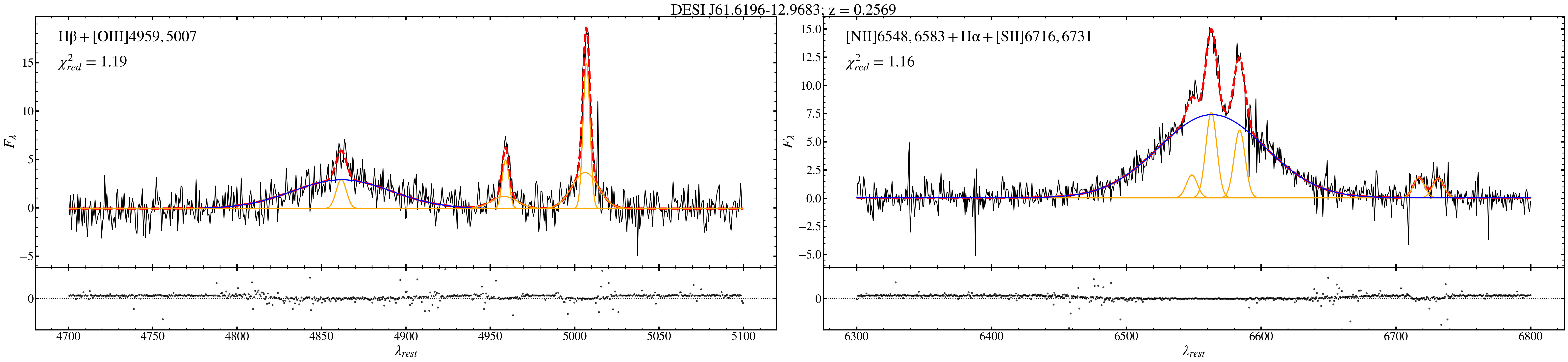}
    \end{subfigure}

\caption{{\bf Example fits of galaxy spectra fit via the EBL mode}, without (top panel) and with (bottom panel) an outflow component in \oiii~emission line. In all the panels, the best-fit models to the continuum-subtracted emission-line spectrum in the regions of $\hb$~+~\oiii~and \nii~+~$\ha$~+~\sii~are shown from left to right. The colors and text in the panels are the same as Figure~\ref{fig:default_broad}.}
    \label{fig:ebl}
\end{figure*}

With just one year of observations, DESI has already obtained spectra of nearly 18 million galaxies and quasars. Within the redshift range relevant to this study (0.001 $\le~z~\le$ 0.45), this sample includes a wide diversity of spectra. {\tt EmFit} is designed to robustly handle this diversity, producing a broad range of reliable fits. In this Section, we present some representative examples of DESI spectra along with their corresponding {\tt EmFit} best-fit models.

{\tt EmFit} operates on two different fitting modes: the ``default'' mode and the ``EBL'' mode. The EBL mode was introduced to accommodate sources whose broad $\ha$ emission extends into the \sii~region, causing the default mode to fail for a subset of them. Of the 7,378,347 galaxies in the VAC, 7,363,400 galaxies are fit with the default mode, and 14,947 galaxies are fit with the EBL mode.

In both modes, emission lines are modeled using one or more Gaussian functions, together with a non-zero polynomial to account for any residual continuum. In the default mode, the \sii~and \oiii~emission lines are independently tested for the presence of any additional components, which may represent either to a second narrow component or to a broader outflow component. If \sii~requires a second component, a corresponding second component is also included while fitting the \nii, $\ha$, and $\hb$ emission lines. The Balmer lines are further tested for the presence of a broad component, irrespective of the extra components in the lines. A full description of the fitting procedure is detailed in \citet{Pucha+2025}.

Figure~\ref{fig:default_nobroad} presents three examples of galaxy spectra fit with the default mode. The top panel shows a narrow-line only galaxy spectrum, in which all lines are well fit by a single Gaussian component. The middle panel illustrates a double-peaked emission-line galaxy, requiring two narrow components per emission line.  The bottom panel shows a case in which the additional detected components correspond to an outflow feature. In the latter two cases, the extra component appears across all emission lines, although this is true only for a subset of galaxies. The criteria for separating double-peaked emission from outflow emission profiles are summarized in Appendix A of \citet{Pucha+2025}. Two columns are added to the VAC using these criteria $-$ {\tt OIII\_DBL\_FLAG} and {\tt SII\_DBL\_FLAG}, which indicate whether the extra component identified in the emission line is a double narrow emission-line peak. In the {\tt EmFit} VAC, we have 18,446 galaxies with {\tt OIII\_DBL\_FLAG = True} and 28,293 galaxies with {\tt SII\_DBL\_FLAG = True}.

Figure~\ref{fig:default_broad} shows three examples of BL candidates fit using the default mode. The top panel shows a galaxy with single narrow components for all emission lines plus broad Balmer components. The middle panel illustrates a BL candidate with double narrow peaks in $\hb$, \nii, and \sii, but only a single narrow component in \oiii~(for this object,  {\tt OIII\_DBL\_FLAG = False} and {\tt SII\_DBL\_FLAG = True}). The bottom panel shows a case where $\hb$, \nii, $\ha$, and \sii~lines require a single narrow component and a broad Balmer component, while \oiii~requires an additional outflow component (for this object, both {\tt OIII\_DBL\_FLAG} and {\tt SII\_DBL\_FLAG = False}). 

Figure~\ref{fig:ebl} presents two examples of BL candidates fit with the EBL mode. The top panel shows a case requiring only broad Balmer components in addition to single narrow peaks for all emission lines. The bottom panel shows an example in which the broad Balmer components are accompanied by an additional outflow component in the \oiii~lines. 

Together, these examples illustrate the flexibility of {\tt EmFit} to model a wide range of narrow- and broad-line galaxy spectra, including cases with complex kinematics and multiple components. 

\begin{figure*}[t!]
    \centering

    \begin{subfigure}{1.0\textwidth}
        \centering
        \includegraphics[width=\textwidth]{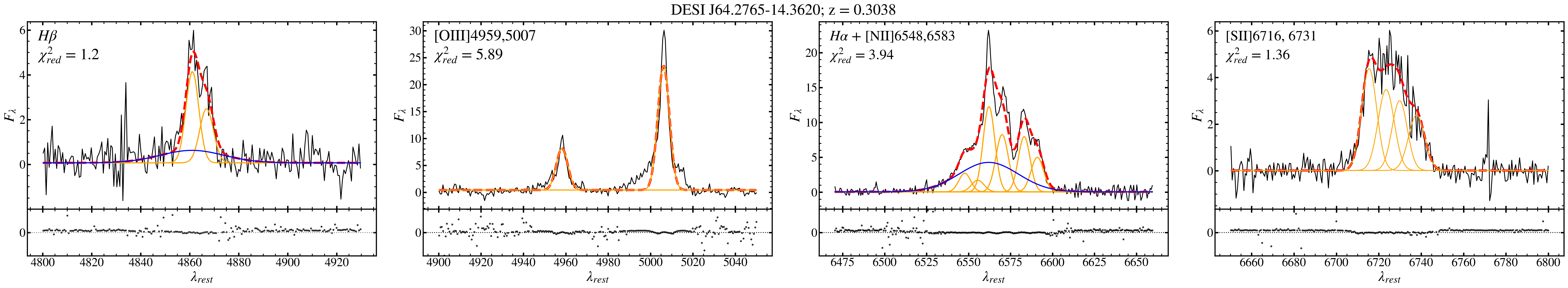}
    \end{subfigure}
    
    \begin{subfigure}{1.0\textwidth}
        \centering
        \includegraphics[width=\textwidth]{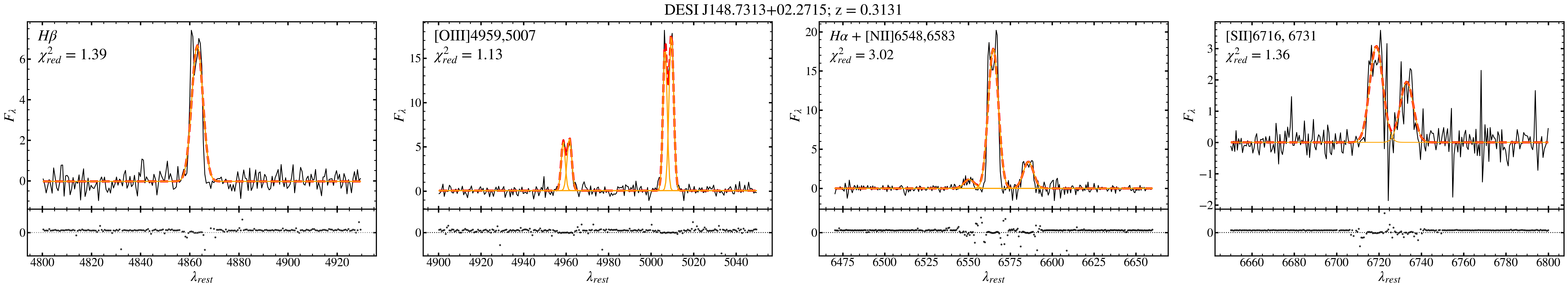}
    \end{subfigure}

    \begin{subfigure}{1.0\textwidth}
        \centering
        \includegraphics[width=\textwidth]{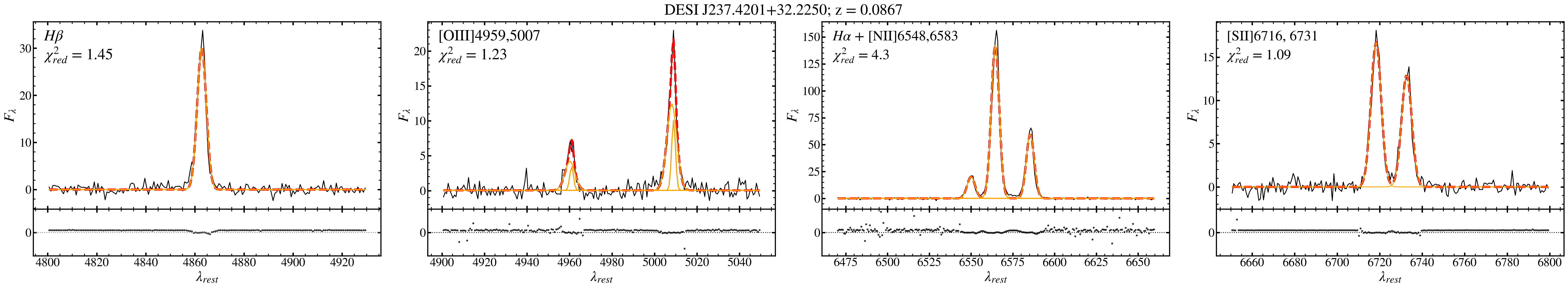}
    \end{subfigure}

    \caption{Example fits of cases with missed outflow (top panel), missed second peak (middle panel), and likely forced extra (bottom panel) components. The descriptions, colors, and texts in the panels are same as Figure~\ref{fig:default_broad}.}
    \label{fig:outflow_issue}
    
\end{figure*}

\begin{figure*}[h!]
    \centering

    \begin{subfigure}{1.0\textwidth}
        \centering
        \includegraphics[width=\textwidth]{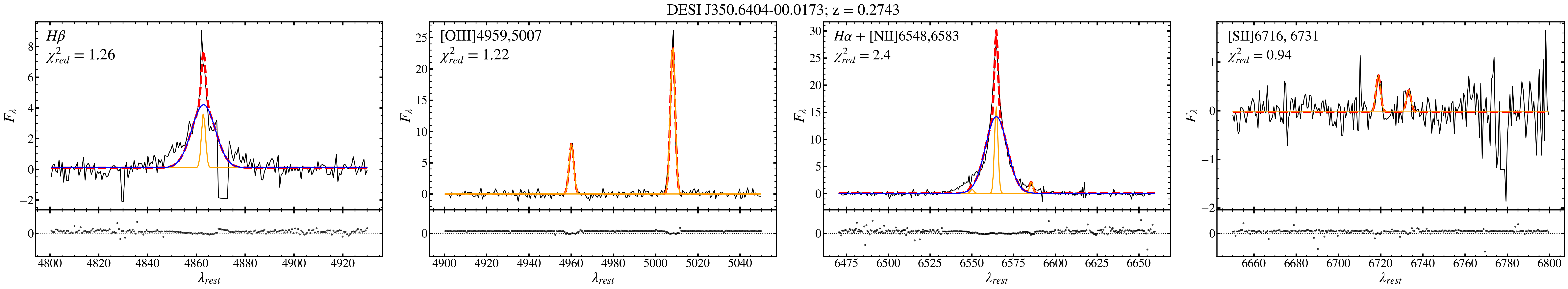}
    \end{subfigure}
    
    \begin{subfigure}{1.0\textwidth}
        \centering
        \includegraphics[width=\textwidth]{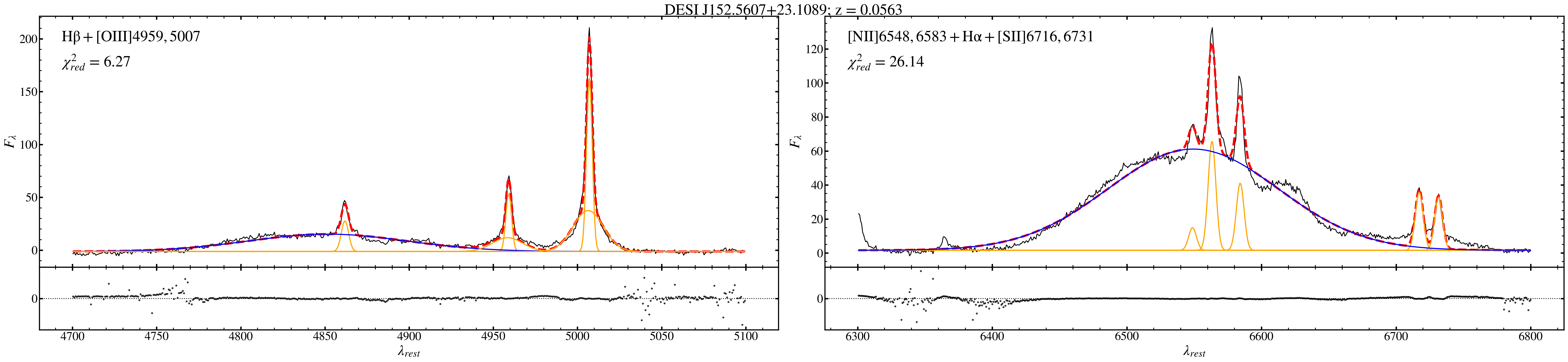}
    \end{subfigure}

    \caption{Examples of cases showcasing the approximation of a single Gaussian for the broad Balmer component. The top panel shows an example fit with default mode ($\hb$, \oiii, \nii~+ $\ha$, and \sii~from left to right), while the bottom panel shows an example fit using EBL mode ($\hb$ + \oiii~and \nii~+ $\ha$ +\sii~from left to right). The colors and texts in the panels are the same as Figure~\ref{fig:default_broad}.}  
    \label{fig:single_gaussian_bl}
\end{figure*}

\begin{figure*}[h!]
    \centering

    \begin{subfigure}{1.0\textwidth}
        \centering
        \includegraphics[width=\textwidth]{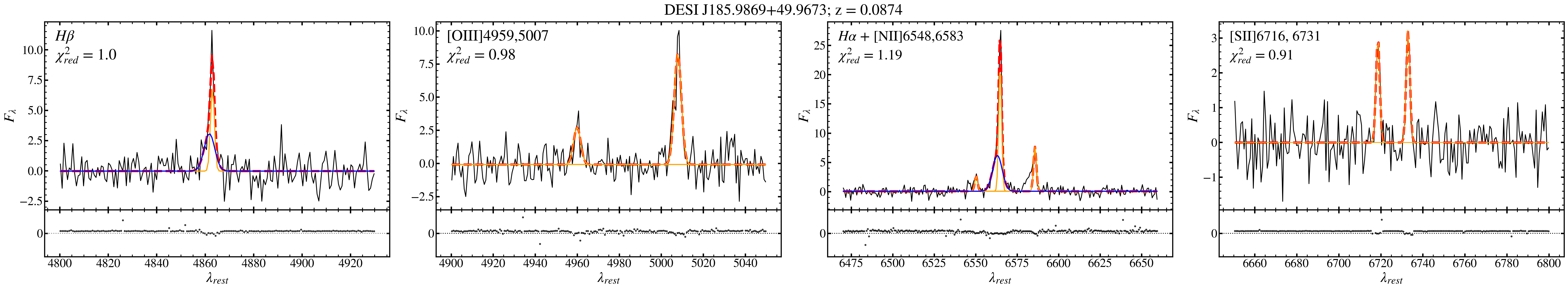}
    \end{subfigure}
    
    \begin{subfigure}{1.0\textwidth}
        \centering
        \includegraphics[width=\textwidth]{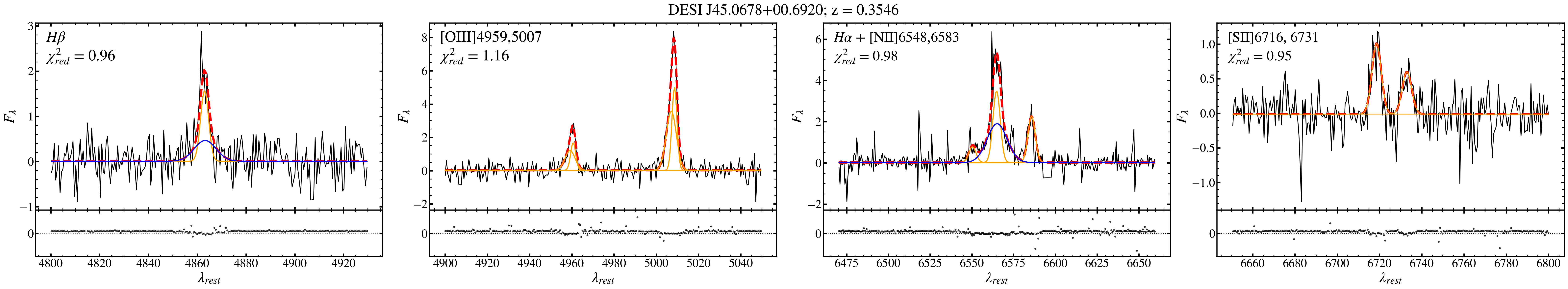}
    \end{subfigure}

    \begin{subfigure}{1.0\textwidth}
        \centering
        \includegraphics[width=\textwidth]{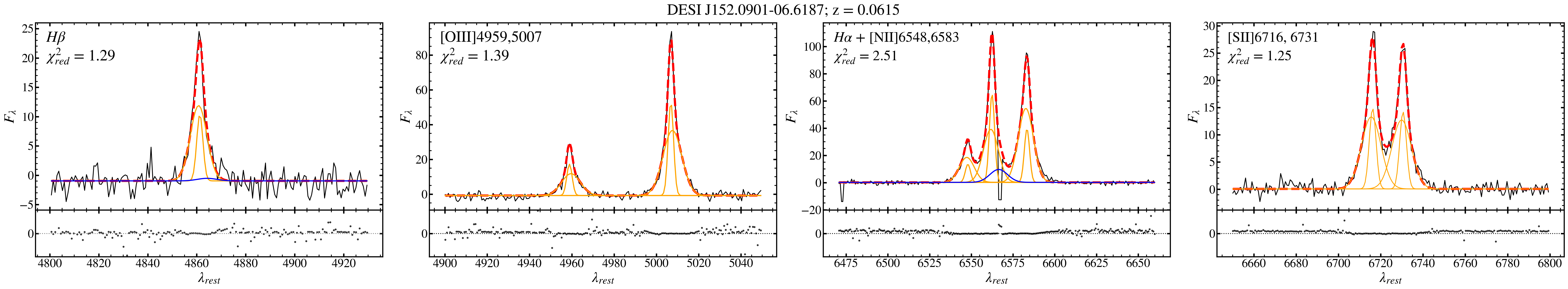}
    \end{subfigure}

    \caption{Example fits of BL candidates with unclear origin of the broad components. The colors and texts in the panels are the same as Figure~\ref{fig:default_broad}.}    
    \label{fig:narrow_broad_lines}
\end{figure*}

\begin{figure*}[h!]
    \centering

    \begin{subfigure}{1.0\textwidth}
        \centering
        \includegraphics[width=\textwidth]{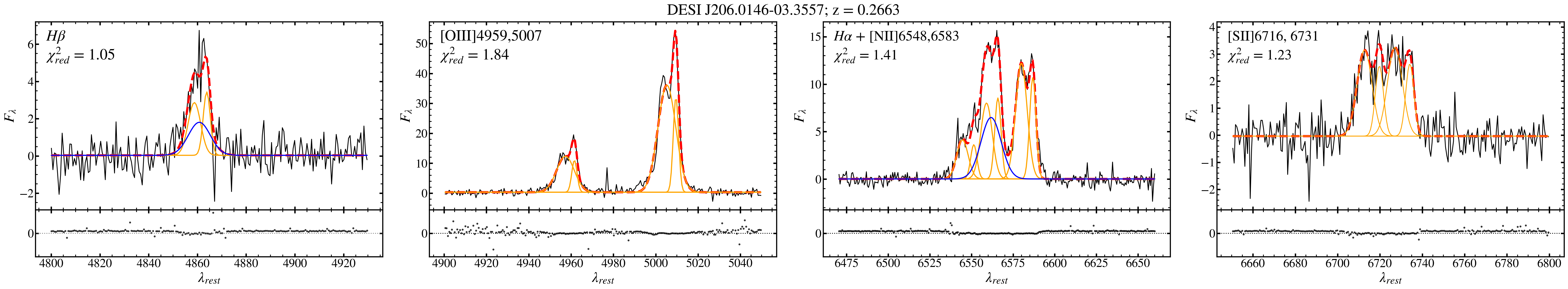}
    \end{subfigure}
    
    \begin{subfigure}{1.0\textwidth}
        \centering
        \includegraphics[width=\textwidth]{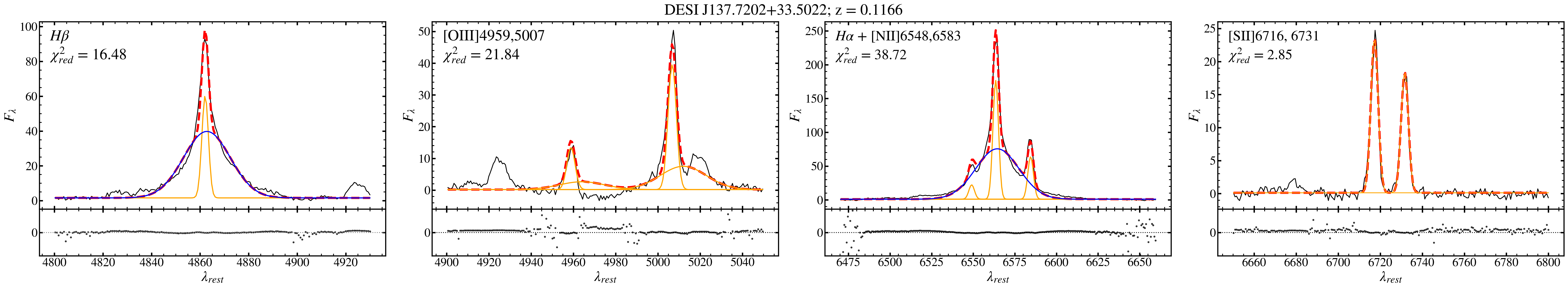}
    \end{subfigure}

    \caption{Examples of complicated fits. The colors and texts in the panels are the same as Figure~\ref{fig:default_broad}.}   
    \label{fig:complicated_fits}
\end{figure*}

\subsection{Known Issues and Caveats \label{app_sub:issues}}

There are some known caveats and limitations associated with {\tt EmFit}, which we summarize below:

\begin{itemize}
    \item As mentioned previously, {\tt EmFit} tests for a second component in the narrow emission lines, which may correspond to either an extra-narrow peak or an outflow component. However, the statistical test used to determine this additional component is not always accurate. In some cases, it fails to detect a visually apparent second component, while in others it introduces a likely spurious component. Figure~\ref{fig:outflow_issue} presents three such examples: the top panel shows a missed \oiii~outflow component which is visually clear; the middle panel shows an example where the second narrow peak is detected only in \oiii, but are missed in the other emission lines; and the bottom panel shows an example where the \oiii~emission line is fit with two narrow components, despite appearing consistent with a single narrow profile. Such cases are also classified as double-peaked emission lines in the {\tt EmFit} VAC, i.e., {\tt *\_DBL\_FLAG = True}.
    
    \item {\tt EmFit} models the broad Balmer component using a single Gaussian component. While this assumption is sufficient for most spectra, it may not capture the more complex profiles observed in some galaxies. Figure~\ref{fig:single_gaussian_bl} shows two examples of BL candidates, fit with default (top panel) and EBL (bottom panel) modes, where the spectra exhibit a more complex profile in the $\ha$ emission line. In such cases, either a non-Gaussian profile is needed, or multiple components may be required \citep[][]{Greene&Ho2005, Liu+2019}. These are beyond the current scope of {\tt EmFit}.
    
    \item Broad Balmer components are accepted when they have $\rm FWHM_{\ha;b} \ge 300~\kms$. This threshold can admit components that do not arise from the virialized gas near the BH. We therefore urge the users to be cautious when interpreting broad components with $\rm FWHM_{\ha;b} < 1000~\kms$. Visual inspection is often the most reliable way to validate these detections. Figure~\ref{fig:narrow_broad_lines} illustrates three examples of such uncertain broad components. The top panel shows an example where a missed outflow as observed in \nii~lines is identified as a broad Balmer component. The middle panel depicts a spectrum with a statistically detected, but visually ambiguous, broad component. The bottom panel shows a complex \nii~+~$\ha$ line profile requiring more careful analysis. More information regarding these is available in Appendix B of \citet{Pucha+2025}.
    
    \item The top panel of Figure~\ref{fig:complicated_fits} shows a visually acceptable fit in which multiple components may originate from complex internal kinematics. Such cases are rare but require caution when interpreting the individual component measurements. 
    
    \item The bottom panel of Figure~\ref{fig:complicated_fits} shows an example in which the local continuum is not properly subtracted near the \oiii~region. This is again a rare case, and they do not significantly affect the recovered emission line fluxes and widths. 

\end{itemize}

Some cautionary notes related to {\tt EmFit} available here: \url{https://data.desi.lbl.gov/doc/releases/dr1/vac/emfit/#notes}.

\section{Selecting Confident BL-AGN Candidates \label{app:vi}}

To construct a reliable $\mbh - \mstar$ scaling relation (Section~\ref{sec:mbh-mstar-relation}), we need to identify confident BL-AGN candidates. As noted in Appendix~\ref{app:emfit}, BL candidates with $\rm FWHM_{\ha;b} \le 1000~\kms$ should be considered with caution. \citet{Pucha+2025} addressed this issue by visually inspecting all such low FWHM candidates, classifying them as confident and tentative BL-AGN candidates. However, in our larger sample of 2,099 extended and 210 point-like BL-AGN candidates, such a visual inspection is not practical. To navigate this, we visually inspect the 210 point sources and perform stacked-spectra analysis for the 2,099 extended sources as described below. 

\begin{figure*}[t!]
    \centering
    \includegraphics[width=\linewidth]{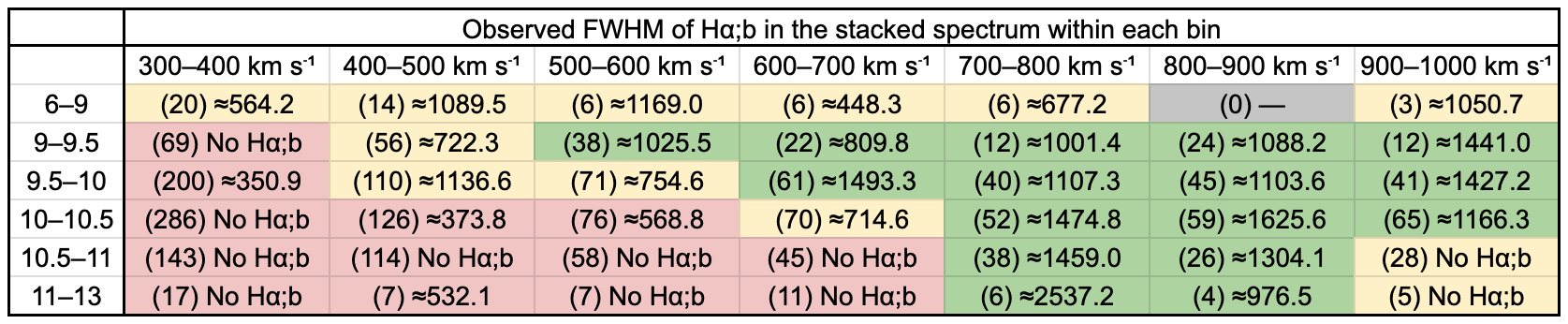}
    \caption{Visual Inspection Results of Spectral Stacking and {\tt EmFit} Fits. {\bf Note.} The first column lists the $\logmass$ bins, while the first row indicates the $\rm FWHM_{\ha;b}$ bins used to group and stack the BL-AGN candidates. Each cell shows the number of galaxies in parentheses in the bin, along with the observed $\rm FWHM_{\ha;b}$ measured from their stacked spectrum. Cells are color-coded based on the visual inspection of the resulting fits: green represents confident broad components, yellow indicates tentative detections, and red marks unreliable candidates.}
    \label{fig:fwhm_logmass}
\end{figure*}

The main motivation for using the stacking process is to boost the signal-to-noise ratio and to perform accurate decomposition of emission line components, including narrow lines, secondary outflow components, and broad permitted lines. We separate the 2,099 extended BL-AGN candidates by stellar masses and $\rm FWHM_{\ha;b}$ using the bins listed in Figure~\ref{fig:fwhm_logmass}. For each bin, we retrieve the DESI spectra of all galaxies from the SPectra Analysis \& Retrievable Catalog Lab \citep[SPARCL;][]{sparcl}. We then employ the functionality from {\tt desigal}\footnote{\url{https://github.com/desihub/desigal}} to perform the following steps. First, we apply the Galactic dust reddening correction based on the E(B-V) value from the \citet{extinction_sfd98} dust map reported in the LS DR9 catalog~\citep[][]{desi_imaging}. Then, we shift the spectra to the rest-frame and resample them on a reference wavelength grid spanning $3350-9800$ \AA~with an increment of 0.6\AA~ensuring flux conservation. We normalize the individual spectra by the mean flux density over $6380-6480$ \AA~before combining them using an inverse-variance weighted mean. We perform error propagation from the individual error spectra to compute the resulting inverse variance on the stacked spectrum. We refit each stacked spectrum using {\tt EmFit} and visually inspect the resulting fits for broad $\ha$ components. 

Figure~\ref{fig:fwhm_logmass} summarizes these results. Each cell lists the number of galaxies in that bin in parentheses and the FWHM of any detected broad component. Cells are color-coded based on the quality of the stacked spectrum fit and our visual inspection:

\begin{itemize}
    \item {\bf Green:} confident broad components in the stacked spectrum $-$ we retain all 545 galaxies in these bins.
    \item {\bf Red:} no broad components or poor quality fit in the stacked spectrum $-$ we remove all 1,159 galaxies within these bins.
    \item {\bf Yellow:} tentative broad components in the stacked spectrum $-$ we visually inspect the 299 galaxies in these bins, resulting in 46 good fits. The remaining objects either show poor fits (11 galaxies) or their broad components are missed outflows \citep[221 galaxies; see Appendix~B of ][]{Pucha+2025}.
\end{itemize}

Of the 210 point BL-AGN candidates, we have 118 sources with reliable broad components. Along with the 46 extended BL-AGN candidates, we include the final sample of 226 low-FWHM BL-AGN candidates for the analysis presented in Section~\ref{sec:mbh-mstar-relation}.

\bibliography{references}{}
\bibliographystyle{aasjournal}

\suppressAffiliationsfalse
\allauthors

\end{document}